\documentclass[conf]{new-aiaa}

\usepackage[colorinlistoftodos]{todonotes} 

\usepackage[utf8]{inputenc}
\usepackage[version=4]{mhchem}
\usepackage{
  amsmath,
  amsfonts,
  bm,
  amsfonts,
  float,
  algorithm,
  algcompatible,
  xcolor,
  graphicx,
  subcaption,
  longtable,
  tabularx,
  siunitx,
  threeparttable,
}
\newcommand{\multiline}[1]{%
  \begin{tabularx}{\dimexpr\linewidth-\ALG@thistlm}[t]{@{}X@{}}
    #1
  \end{tabularx}
}
\newcommand{\range}[1]{0, \dots, #1}
\newcommand{\rangeset}[1]{\{\range{#1}\}}
\newcommand{\andc}{~~\text{and}~~}
\newcommand{\atantwo}[2]{\text{atan2}(#1, #2)}
\title{Planning Visual Inspection Tours for a 3D Dubins Airplane Model in an Urban Environment}
\author{Collin Hague \footnote{Graduate student, Department of Mechanical Engineering and Engineering Science}, Andrew Willis \footnote{Associate Professor, Department of Electrical and Computer Engineering}, Dipankar Maity \footnote{Assistant Professor, Department of Electrical and Computer Engineering}, Artur Wolek \footnote{Assistant Professor, Department of Mechanical Engineering and Engineering Science, Member AIAA}}
\affil{University of North Carolina at Charlotte, Charlotte, North Carolina, 28223}
\begin{document}
\maketitle

\begin{abstract}
This paper investigates the problem of planning  a minimum-length tour for a three-dimensional Dubins airplane model to visually inspect a series of targets located on the ground or exterior surface of objects in an urban environment. Objects are 2.5D extruded polygons representing buildings or other structures. A visibility volume defines the set of admissible (occlusion-free) viewing locations for each target that satisfy feasible airspace and imaging constraints. The Dubins traveling salesperson problem with neighborhoods (DTSPN) is extended to three dimensions with  visibility volumes that are approximated by triangular meshes. Four sampling algorithms are proposed for sampling vehicle configurations within each visibility volume to define vertices of the underlying DTSPN. Additionally, a heuristic approach is proposed to improve computation time by approximating edge costs of the 3D Dubins airplane with a lower bound that is used to solve for a sequence of viewing locations. The viewing locations are then assigned pitch and heading angles based on their relative geometry. 
The proposed sampling methods and heuristics are compared through a Monte-Carlo experiment that simulates view planning tours over a realistic urban environment.
\end{abstract}


\section{Introduction}

\lettrine{U}{nmanned} aerial vehicles (UAVs) are routinely used in applications such as visual reconnaissance, infrastructure inspection, and aerial photography to image a series of points of interest (henceforth referred to   as  \emph{targets}). 
In three-dimensional environments (e.g., an urban city,  mountainous terrain) the targets must be imaged from particular vantage points to avoid occlusions from surrounding objects (e.g., buildings, trees). Additional requirements, such as airspace restrictions and image resolution, further constrain the three-dimensional \emph{visibility volume} from which an image of a target may be obtained. This paper investigates the problem of planning a path to image a set of targets by flying through their corresponding visibility volumes  in minimum time. The UAV is modeled as a \emph{Dubins airplane} \cite{chitsaz2007time, ambrosino2009path} and the environment consists of extruded polygonal objects with targets located on the ground or on the surface of objects.

\subsection{Relation to Prior Work}
The view planning problem considered here is related to the \emph{Dubins traveling salesperson problem} (DTSP \cite{NyEtAl.IEEE.2012}) of constructing a minimum-time tour for a constant-speed planar \emph{Dubins vehicle} model \cite{dubins1957curves} to travel through a series of planar points (with arbitrary heading).  The set of points to visit can be generalized to arbitrary planar regions (e.g., polygons) to give the \emph{DTSP with neighborhoods} (DTSPN \cite{Isaacs.ACC.2011}) wherein the Dubins vehicle must visit at least one point in each region/neighborhood. One application of the DTSPN is to plan visual inspection tours for an airplane to visit planar polygonal regions at a constant altitude to image ground targets  \cite{Obermeyer2012}.
More recently, the \emph{Dubins airplane} model \cite{chitsaz2007time, ambrosino2009path} that includes additional degrees of freedom (altitude and pitch angle) was used to extend the DTSPN to three dimensions. Planning three-dimensional Dubins tours have typically assumed that the desired viewing regions have relatively simple geometries, such as spheres \cite{faigl2018surveillance} or cylinders \cite{vavna2018dubins}. In contrast, this work admits more complex target visibility volumes that are approximated as triangular meshes. 
\subsection{Contributions}
This paper formulates a view planning problem for a 3D Dubins airplane model to observe a set of targets occluded by objects in an urban environment.
The contributions of the paper are: (1) four sampling algorithms that extend two-dimensional Dubins-based view planning to three dimensions with visibility volumes that have an arbitrary geometry approximated by a triangular mesh, and (2) a heuristic approach that solves for a tour using a modified Euclidean distance TSP (METSP) with edge costs that are lower bounds for the 3D Dubins path length and using the geometry of consecutive viewing locations in the METSP tour to  assign heading and pitch angles. The relative performance of the algorithms are characterized through a Monte-Carlo experiment. 

\subsection{Paper Organization}
The remainder of the paper is organized as follows. Section~\ref{sec:problemStatement} describes the airplane motion model, the environment model, the target visibility volumes, and states the view planning problem.
Section~\ref{sec:baseline} describes a method for  approximately computing the target visibility volumes and path planning for constant-altitude 2D tours. Section~\ref{sec:proposed} introduces 3D path planning algorithms and proposes heuristics to reduce computation time. Section~\ref{sec:results} describes the results of a Monte-Carlo experiment that compares the 2D and 3D algorithms. The paper is concluded in Sec.~\ref{sec:conclusion}.

\section{Problem Formulation}
\label{sec:problemStatement}
This section formulates the problem of planning a minimum time path for an unmanned airplane to visually inspect a set of targets in the presence of occluding structures. The vehicle motion model, environmental model, and target visibility volumes are introduced, and the view planning problem is formally stated.
\subsection{Airplane Motion Model}
\begin{figure}[h!]
  \centering
  \includegraphics[width=0.5\textwidth]{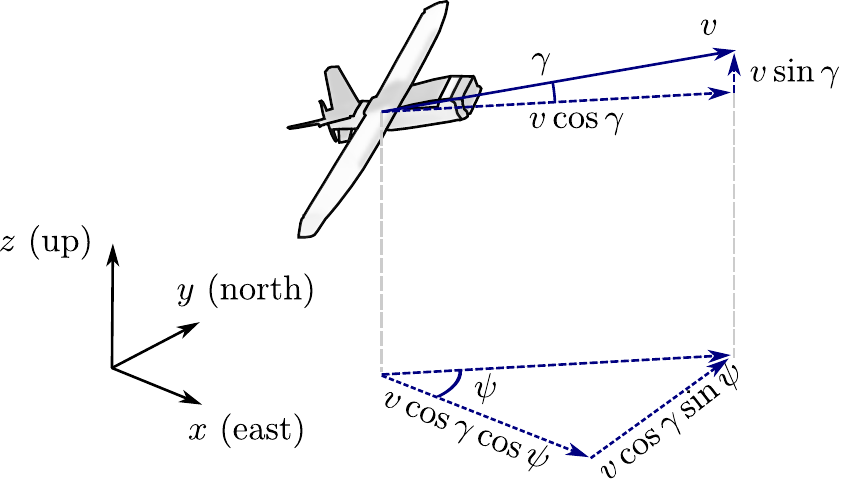}
  \caption{The model for a Dubins airplane flying at speed $v$ where $(x,y,z)$ is the inertial position, $\psi$ is the heading angle, and $\gamma$ is the pitch angle.}
  \label{fig:model}
\end{figure}

\begin{figure}[h!]
  \centering
  \includegraphics[scale=.6]{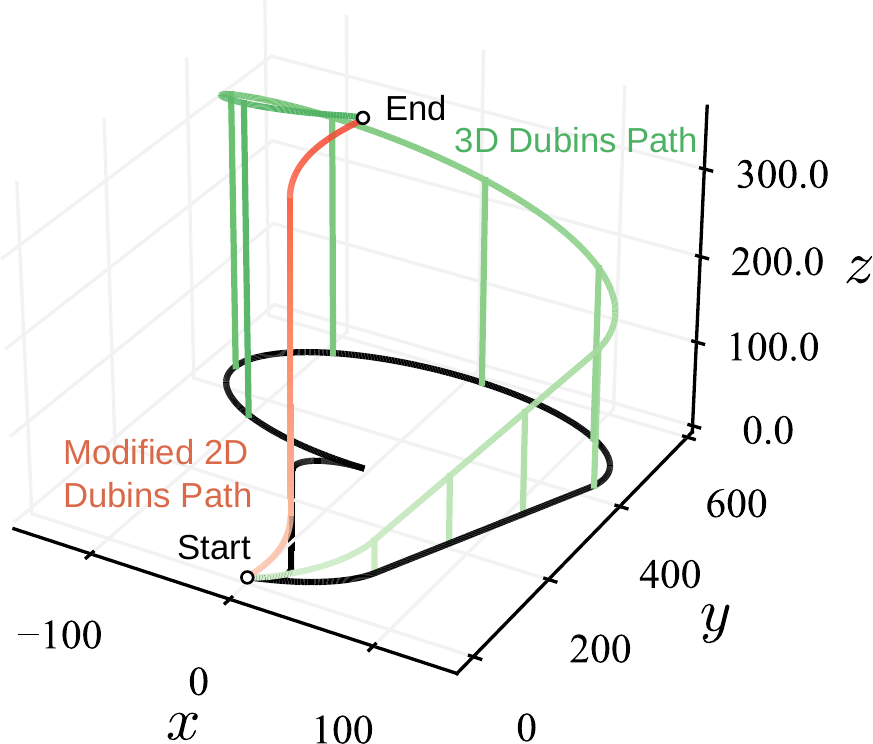}
  \caption{
    An example 3D Dubins airplane path (green) \cite{Vana2020} joining configurations $\bm{q}_1=(0,0,0,\frac{\pi}{6},0)$ and $\bm{q}_2=(0, 300~\rm{m}, 400~\rm{m}, 0, 0)$ is compared to a modified 2D Dubins path (red) that join the same pair of locations and heading angles. The modified 2D Dubins path is shorter (523 m compared to 1184 m) but violates the pitch angle constraint since a large altitude change is required over a relatively short distance. The paths are constructed with the parameters: $\rho_{\rm min}=40~\rm{m}$, $\gamma_{\rm min} = {-\pi}/{12}$, and  $\gamma_{\rm max}= {\pi}/{9}$.
  }
  \label{fig:dubins}
\end{figure}

This work considers the three-dimensional Dubins airplane model \cite{Owen2015, Vana2020}:
\begin{gather}
\begin{bmatrix}
\dot{x}\\\dot{y}\\\dot{z} \\ \dot{\psi} \\ \dot{\gamma}
\end{bmatrix}
=
\begin{bmatrix}
v\cos{\psi}\cos{\gamma}\\
v\sin{\psi}\cos{\gamma}\\
v\sin{\gamma} \\
u_\psi \\
u_\gamma 
\end{bmatrix}
\;,
\end{gather}
where $(x,y,z) \in \mathbb{R}^3$ is the inertial position of the airplane expressed in an east-north-up coordinate system, $v$ is the vehicle's speed, $\psi$ is the heading angle, and $\gamma$ is the pitch angle (see Fig.~\ref{fig:model}).
The control inputs are the turn-rate $u_\psi$ and the pitch-angle-rate $u_\gamma$. 
The Dubins airplane model travels in the direction it is pointed so that the pitch angle $\gamma$ is equivalent to the flight path angle and is constrained between a minimum and maximum angle, $\gamma \in [\gamma_{\rm min}, \gamma_{\rm max}]$. The controls are constrained such that the path curvature $\rho_{\min}$ is bounded \cite{Wang2015}:
\begin{equation}
    \rho_{\min} \leq \frac{1}{\sqrt{u_\psi^2 \cos^2{\gamma} + u_\gamma^2}}\;.
    \label{eqn:radiusBound}
\end{equation}
Let the vehicle's configuration be denoted $\bm{q}=(x,y,z,\psi,\gamma) \in Q$ where $Q=\mathbb{R}^3 \times \mathbb{S}^2$ is the configuration space.
An example 2D Dubins path (modified with a constant pitch angle to join two altitudes) and a 3D Dubins path that join $\bm{q}_i = (x_j, y_j, z_j, \psi_j, \gamma_j)$ and $\bm{q}_j = (x_j, y_j, z_j, \psi_j, \gamma_j)$ are  shown in Fig.~\ref{fig:dubins}.
The modified 2D Dubins path uses a constant pitch angle $\gamma_c$ that is computed from the change in altitude and planar displacement between the start and end configurations.
The modified 2D Dubins path does not satisfy the required pitch angle at the start/end configurations and may violate pitch angle constraints along the path when the change in altitude is large relative to the planar displacement.
Instead, a 3D Dubins path can join two configurations  while limiting the pitch angle along the path to within the allowable bounds.
The 3D Dubins paths are generated according to \cite{Vana2020} by decomposing the 3D path into two decoupled 2D Dubins paths.
First, a 2D horizontal Dubins path is constructed in the $xy$ plane to join the 2D Dubins configurations $(x_i, y_i, \psi_i)$ and $(x_j, y_j, \psi_j)$ using a horizontal turn radius that is twice the minimum turn radius $\rho_{\rm h}=2\rho_{\rm min}$.
Next, a 2D vertical path is constructed, with vertical plane turn radius $\rho_{\rm v}$ that is found from \cite{Vana2020}
\begin{equation}
    \rho_{\rm min}^{-2} = \rho_{\rm h}^{-2} + \rho_{\rm v}^{-2}\;,
    \label{eqn:radius}
\end{equation}
to join the 2D Dubins configurations $(s_i, z_i, \gamma_i)$ and $(s_j, z_j, \gamma_j)$ where $s_i$ and $s_j$ are the initial and final arc-lengths along the Dubins path in the $xy$ plane (where $s_i = 0$).
The turn radii, $\rho_{\rm h}$ and $\rho_{\rm v}$, are iteratively varied while satisfying \eqref{eqn:radius} to meet the acceptable pitch angle constraint while minimizing the path length as described in \cite{Vana2020}.
The length of a 3D Dubins path between two configurations, $\bm{q}_i, \bm{q}_j \in {Q}$ is denoted $D(\bm{q}_i, \bm{q}_j): {Q}^2 \rightarrow \mathbb{R}$.
\subsection{Environment}
\label{sec:environment}
The airplane operates in an urban environment that consists of a ground plane and a collection of 2.5-dimensional objects representing buildings or other structures.
Let $O = \{O_0,\ldots,O_{N_O -1}\}$ be the set of $N_O$ objects, where $O_i \subset \mathbb{R}^3$ for each $i \in \rangeset{N_O - 1}$.
The $i$th object is an extruded polygon $O_i = \{(x,y,z) \in \mathbb{R}^3~|~(x,y) \in A_i$ and $z \in [0, h_i]\}$ where $A_i \subset{R}^2$ is the object's footprint and $h_i$ is the height of the object.
The set of points along the boundary of $A_i$ is a simple two-dimensional polygon denoted $\partial A_i$ whose shape is defined by an ordered set of points with a positive signed area.
Points on the interior of $A_i$ belong to the set  denoted ${\rm int}(A_i)$. 
The polygonal areas of each object do not intersect ${\rm int}(A_i) \cap {\rm int}(A_j) = \emptyset$ for all $i\neq j$ with $i,j \in \rangeset{N_O-1}$.
The height of the tallest object in $O$ is denoted $h_{\rm max}$, and the airplane is constrained to fly in a feasible airspace 
\begin{equation}
    F=D\times [z_{\rm min}, z_{\rm max}] - O\;,
    \label{eq:feasible_airspace}
\end{equation} 
where $D \subset \mathbb{R}^2$ is the planar region containing the polygonal objects, i.e., $A_i \subset D$ for all $i\in\rangeset{N_O-1}$, $z_{\rm min}$ and $z_{\rm max} > z_{\rm min}$ are the minimum and maximum operating altitudes of the airplane.
The union of all the objects is subtracted from the rectangular volume $D \times [z_{\rm min}, z_{\rm max}]$ in \eqref{eq:feasible_airspace}. 
To ensure that 3D Dubins paths joining two configurations does not exceed the feasible airspace or encounter obstacles, the feasible airspace and set of objects can be artificially contracted and inflated, respectively.
This work assumes that the minimum altitude $z_{\rm min}$ is constrained to be above the tallest building, $z_{\rm min} > h_{\rm max}+ 2\rho_{\rm min}$, such that the airplane's feasible airspace is free of objects and there is enough vertical space to maneuver without collision. 
\subsection{Target Visibility Volumes}
\label{sec:vis_vols}
The airplane is assumed to be equipped with a gimbaled camera and is tasked with inspecting a set of $M$ targets located at the points $P = \{\bm{p}_0,\dots, \bm{p}_{M-1}\}$. 
Each target ${\bm p} = (p_{x}, p_{y}, p_{z}) \in P$ is located in an unobstructed area of the ground plane or on the exposed surface of an object.
That is, each target has planar location $(p_{x}, p_{y}) \in D$ and altitude $p_{z}$ satisfying the following cases: (i) if $(p_{x}, p_{y}) \cap A_i = \emptyset$ for all $i \in \rangeset{N_O-1}$ then the target is on the ground plane with $p_{z} = 0$, (ii) if $p_{y} \cap \partial A_{i} \neq \emptyset$ for some $i \in \rangeset{N_O-1}$ then the target is located on the vertical wall of the $i$th object and $p_{z} \in [0, h_i]$, or (iii) if $p_{y} \cap {\rm int}(A_i) \neq \emptyset$  then $p_{z} = h_i$ such that the target is on top of the $i$th object. 
For each target, a target visibility volume $V_i$ is defined as the set of points ${\bm g} \in \mathbb{R}^3$ that have a direct line-of-sight to the target (i.e., not obscured by buildings). Let 
\begin{equation}
    L(\tau; {\bm g}, {\bm p}) = (\bm{p} - \bm{g})\tau + \bm{g}~~\text{for}~~\tau \in [0,1]\;
\end{equation}
denote a line segment that joints two points ${\bm g},{\bm p} \in \mathbb{R}^3$ where $\tau$ is a normalized arc-length.
The visibility volume for a target located at ${\bm p} = (p_x, p_y, p_z)$ is the subset of the feasible airspace that is within direct line-of-sight to the target, within a maximum range $d_{\rm max}$ relative to the target, and at least a distance $h_{\rm view}$ above the target: 
\begin{equation}
    \begin{split}
        V({\bm p}; F, O, d_{\rm max}, h_{\rm view}) &= 
        \{ {\bm g}=(g_x, g_y, g_z) \in F ~\text{such that}~
        ||{\bm p} - {\bm g} || \leq d_{\rm max},~~
        h_{\rm view} + p_z \leq g_z \andc \\
        & \qquad 
         L(\tau; {\bm g}, {\bm p}) \cap O_j = \emptyset~\text{for all}~\tau \in[0,1]~\text{and}~j \in \rangeset{N_O-1} \}\;.
    \end{split}
    \label{eq:visibility_volumes}
\end{equation}
For brevity, visibility volumes \eqref{eq:visibility_volumes} are henceforth denoted $V({\bm p})$. 
The maximum range $d_{\rm max}$ constraint models minimum image resolution requirements. The minimum height-above-target $h_{\rm view} < d_{\rm max} $ constraint ensures images are captured with sufficient surrounding context (e.g., the point target may actually represent an extended body that should be contained in the image) or to reduce gimbal pointing speed and precision requirements. For the problem to be well posed, there should always exist at least one valid viewing point above each target. This condition may be satisfied by the following parameter constraints:  
\begin{align}
z_{\rm min} &\leq h_{\rm view} + h_{\rm max} \leq  z_{\rm max}\;,  \label{eq:constr1} \\
z_{\rm min} &\leq d_{\rm max}  \label{eq:constr2} \;,  \\
 2d_{\rm max} &< ||{\bm p}_i - {\bm p}_j|| \quad\text{for all}~ {\bm p}_i,{\bm p}_i \in P~\text{with}~{\bm p}_i \neq {\bm p}_j  \label{eq:constr3} \;. 
\end{align}
If a target is located on top of the highest object, then constraint \eqref{eq:constr1} ensures that a viewing point exists that is below the maximum feasible altitude and above the minimum feasible altitude. For targets that are located on the ground plane, constraint \eqref{eq:constr2} ensures that the sensor range is sufficiently large to view the target from the minimum feasible altitude. Lastly, constraint \eqref{eq:constr3} is a simplifying assumption that guarantees targets are spaced sufficiently far apart such that their visibility volumes do not intersect $V({\bm p}_i) \cap V({\bm p}_j) = \emptyset$ for all $i,j \in \rangeset{M-1}$ with $i \neq j$. 
\subsection{View-planning Problem Statement}
Let $B(\bm{q})$ be a mapping from a configuration $\bm{q} = (x, y, z, \psi, \gamma) \in Q$ to an integer in the set $\rangeset{M-1}$ that identifies the visibility volume corresponding to ${\bm q}$, i.e., the integer $B(\bm{q})$ corresponds to the target ${\bm p}_{B(\bm{q})}\in P$ for which
$(x,y,z) \in V({\bm p}_{B(\bm{q})})$. If $\bm{q}$ is not contained in any visibility volume then $B(\bm{q}) = \emptyset$. 
The optimization problem is to find the sequence of vehicle configurations ${{\bm q}_0,\dots,{\bm q}_{M-1}}$ that 
\begin{equation}
{\rm minimize}\qquad \sum_{i=0}^{M-1} D(\bm{q}_i, \bm{q}_{i+1}) + D(\bm{q}_{M-1}, \bm{q}_0)\;,
\label{eq:cost}
\end{equation}
subject to
\begin{equation}
B(\bm{q}_i)\neq B(\bm{q}_j),\quad\text{for all}~i,j \in \rangeset{M-1}~\text{with}~~i\neq j\;,
\label{eq:visitOnce}
\end{equation}
 \begin{equation}
 B(\bm{q}_0)\cup\dots\cup B(\bm{q}_{M-1})=\rangeset{M-1}\;,
 \label{eq:visitAll}
 \end{equation}
where the cost function (\ref{eq:cost}) is the total length of the 3D Dubins paths in the tour, the constraint (\ref{eq:visitOnce} ensures that each vehicle configuration lies within a unique visibility volume and the constraint \ref{eq:visitAll}) ensures that all visibility volumes are visited. The view planning problem \eqref{eq:cost}--\eqref{eq:visitAll} is a mixed continuous/combinatorial optimization problem with a nonlinear cost function and constraints. Since the vehicle travels at a constant speed the minimum-length tour is also the minimum-time tour.
\section{2D Algorithms}
\label{sec:baseline}
In this section, a target visibility volume mesh approximation is described (Sec.~\ref{sec:volumeApproximation}) followed by a description of two-dimensional algorithms  (Sec.~\ref{sec:baselineDTSP} and Sec.~\ref{sec:baselineDTSPN}) that solve the view planning problem \eqref{eq:cost}--\eqref{eq:visitAll}. The algorithms discussed here include (i) traveling directly over each target (i.e., formulating a Dubins traveling salesperson problem (DTSP) \cite{Savla2008}), and (ii) the DTSP with neighborhoods (DTSPN) to visit one point in a set of visibility polygons corresponding to the targets \cite{Obermeyer2012} that is modified to use an optimized altitude for defining the visibility polygons
\subsection{Target Visibility Volume Approximation}
\label{sec:volumeApproximation}
\begin{figure}[h]
  \centering
  \includegraphics[scale=.6]{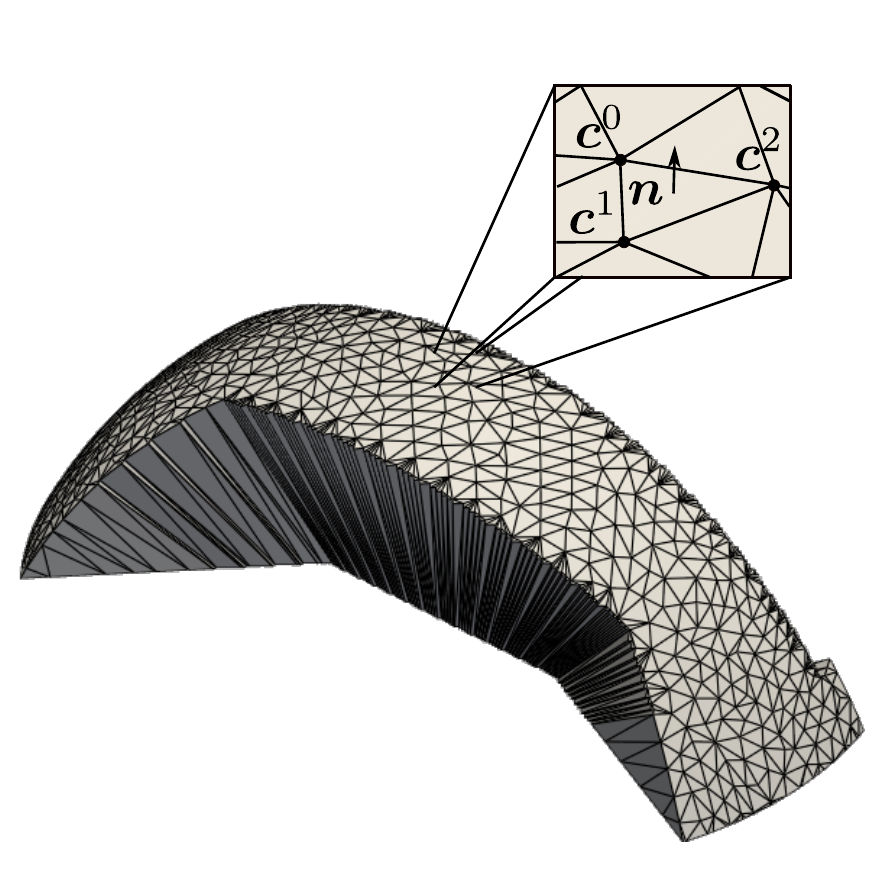}
  \caption{Example target visibility region with a mesh element defined by three vertices $\bm{c}^0$, $\bm{c}^1$, $\bm{c}^2$ and outward pointing normal vector $\bm{n}$.}
  \label{fig:meshElement}
\end{figure}

Volumes in 3D are commonly approximated by a triangular mesh \cite{botsch:inria-00538098}.
While many prior works on the DTSP have assumed simplified 3D geometries (e.g., spheres, cylinders), we propose to use triangular meshes since they can represent arbitrary geometries. The $i$th target visibility volume $V_i$ is approximated with $N_F$ triangular mesh elements resulting in the mesh $\hat{V}_i$. Let $|\hat V_i|$ denote the total number of mesh elements.
The $j$th mesh element in $\hat V_i$ is defined as a set of vectors $\hat V_{ij} = \{ {\bm c}_{ij}^0, {\bm c}_{ij}^1, {\bm c}_{ij}^2, {\bm n}_{ij} \}$ where the vectors ${\bm c}_{ij}^0, {\bm c}_{ij}^1, {\bm c}_{ij}^2 \in \mathbb{R}^3$ are the positions of the vertices of a  triangular mesh element, and ${\bm n}_{ij} \in \mathbb{R}^3$ is an outward pointing normal vector, as illustrated in Fig.~\ref{fig:meshElement}. The mesh-based target visibility volumes $\hat V_i$ are computed using the painter's algorithm \cite{deBerg2008}. A sphere centered on each target is decomposed into six mutually perpendicular views, and each view looks out from the target point location with a 90-degree field-of-view thereby covering one of the six sides of a cube enclosing the point. OpenGL \cite{OpenGL} and a special version of the geometric depth map, i.e., inverse depth,  is used to capture the depth of scene objects in the direction of each view. After calculating the depth values, those that are  less than or equal to $d_{\rm max}$ are tessellated into a preliminary 3D visibility volume mesh. This mesh is genus-0 \cite{botsch:inria-00538098}, i.e., a deformation of the sphere, and is also a manifold surface amenable to constructive solid geometry (CSG) Boolean operations.
Next, the mesh is intersected with the feasible airspace $F$ and the minimum viewing distance constraint $h_{\rm min}$ is imposed 
using CSG Boolean intersection operations.
To reduce the number of vertices in the resulting mesh a decimation procedure is applied  \cite{botsch:inria-00538098}.
\subsection{Baseline Algorithm: Dubins Traveling Salesperson Problem (DTSP)}
\label{sec:baselineDTSP}
The DTSP is the problem of finding the shortest planar tour that visits all points in a graph once using points that are connected with 2D Dubins paths.
Since the objects considered here are extruded polygons, there are no features that can block viewing targets from above (e.g., bridges or tunnels are not admissible). 
Consequently, the view planning problem \eqref{eq:cost}--\eqref{eq:visitAll} can be solved with the DTSP by flying at a fixed altitude directly over each target. All feasible altitudes (i.e., that are common to all visibility volumes) lead to identical cost tours.
To account for the different possible heading angles at each overhead location the heading-angle-discretized DTSP is adopted  \cite{MedeirosUrrutia.2010}.
An example solution is shown in  Fig.~(\ref{fig:dtsp}a).
\subsection{Optimized Altitude DTSP with Neighborhoods (DTSPN)}
\label{sec:baselineDTSPN}
\begin{figure}[h]
    \begin{subfigure}{0.48\textwidth}
      \centering
      \includegraphics[width=0.6\textwidth]{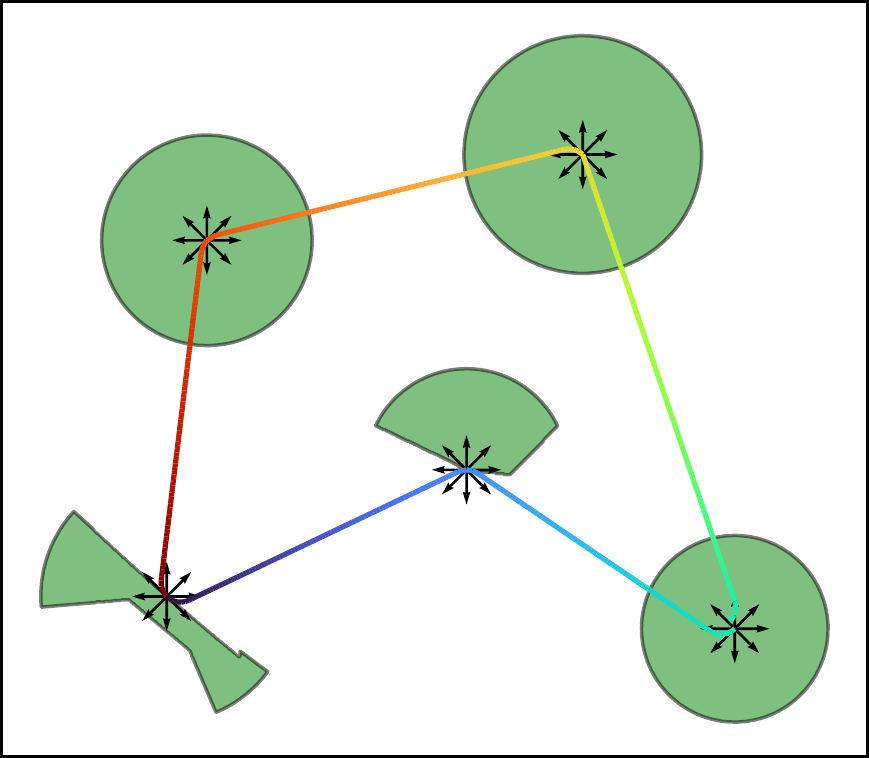}
      \subcaption{2D Dubins traveling salesperson problem (DTSP)}
    \end{subfigure}
    \hfill
    \begin{subfigure}{0.48\textwidth}
      \centering
      \includegraphics[width=0.6\textwidth]{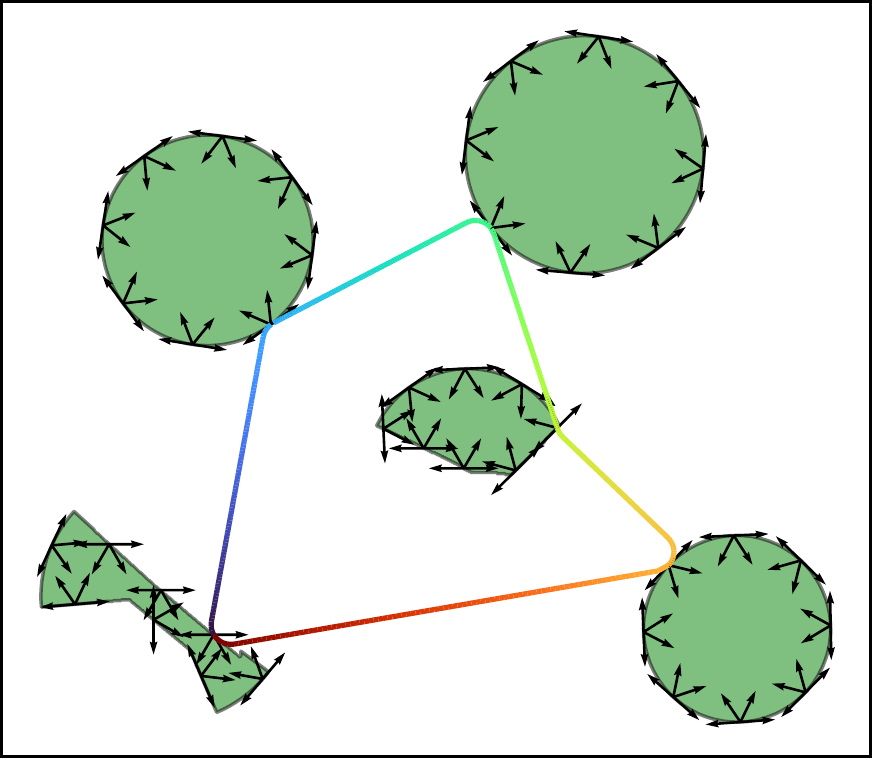}
      \subcaption{2D DTSP with neighborhoods (DTSPN)}
    \end{subfigure}
  \centering
  \caption{
    Example solutions to the view-planning problem using 2D constant-altitude algorithms.
    The green regions are visibility polygons for a chosen altitude $z^*$ while the black arrows represent the heading angle at sampling points.
    The multicolored line is the solution path of the TSPs with  increasing path length represented by color changes from red to purple. The DTSP (a) is solved with entry pose sampling using eight headings samples directly over the targets, while the 2D DTSPN (b) uses eight sample locations around the perimeter of the polygon with four heading angles that are tangent or point into the corresponding visibility polygons.
  }
  \label{fig:dtsp}
\end{figure}

\begin{figure}[ht]
    \begin{subfigure}{0.48\textwidth}
      \centering
      \includegraphics[width=0.6\textwidth]{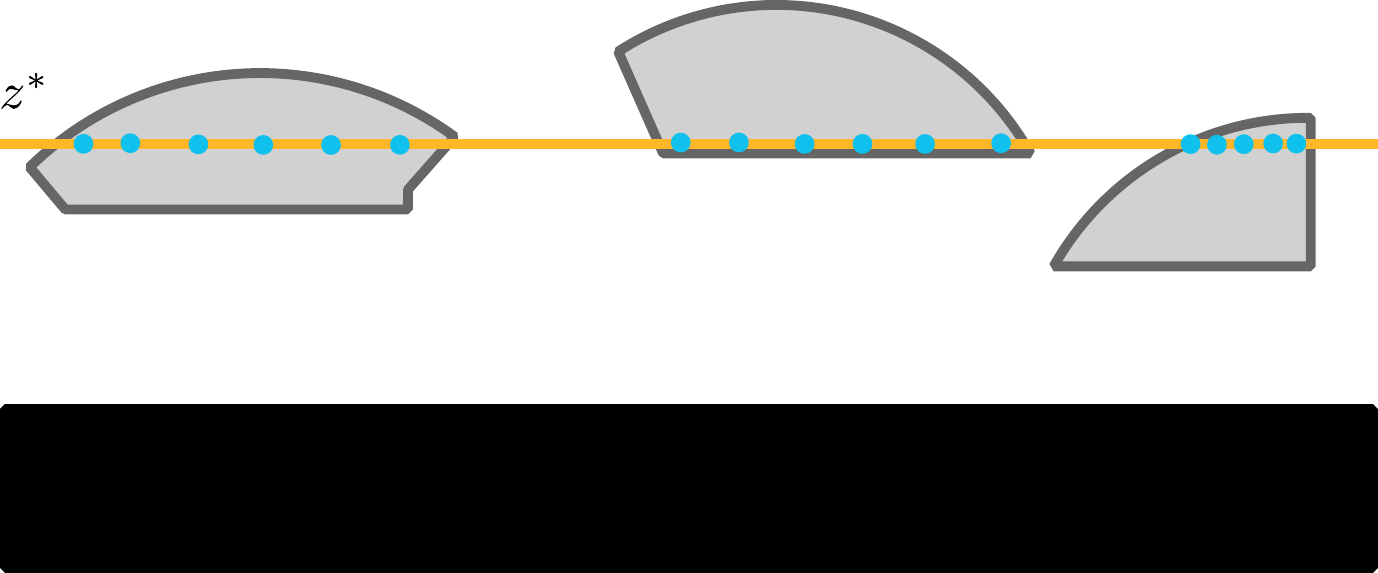}
      \subcaption{Optimized altitude with entry pose sampling}
    \end{subfigure}
    \hfill
    \begin{subfigure}{0.48\textwidth}
      \centering
      \includegraphics[width=0.6\textwidth]{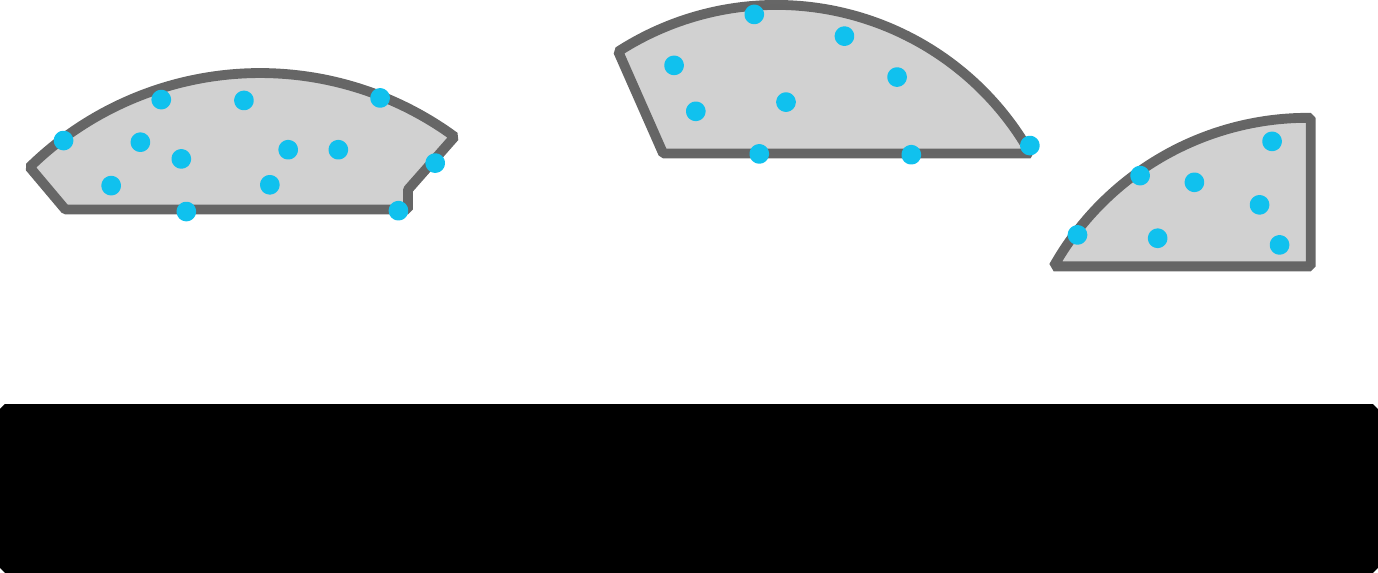}
      \subcaption{Random face sampling}
    \end{subfigure}
    \begin{subfigure}{0.48\textwidth}
      \centering
      \includegraphics[width=0.6\textwidth]{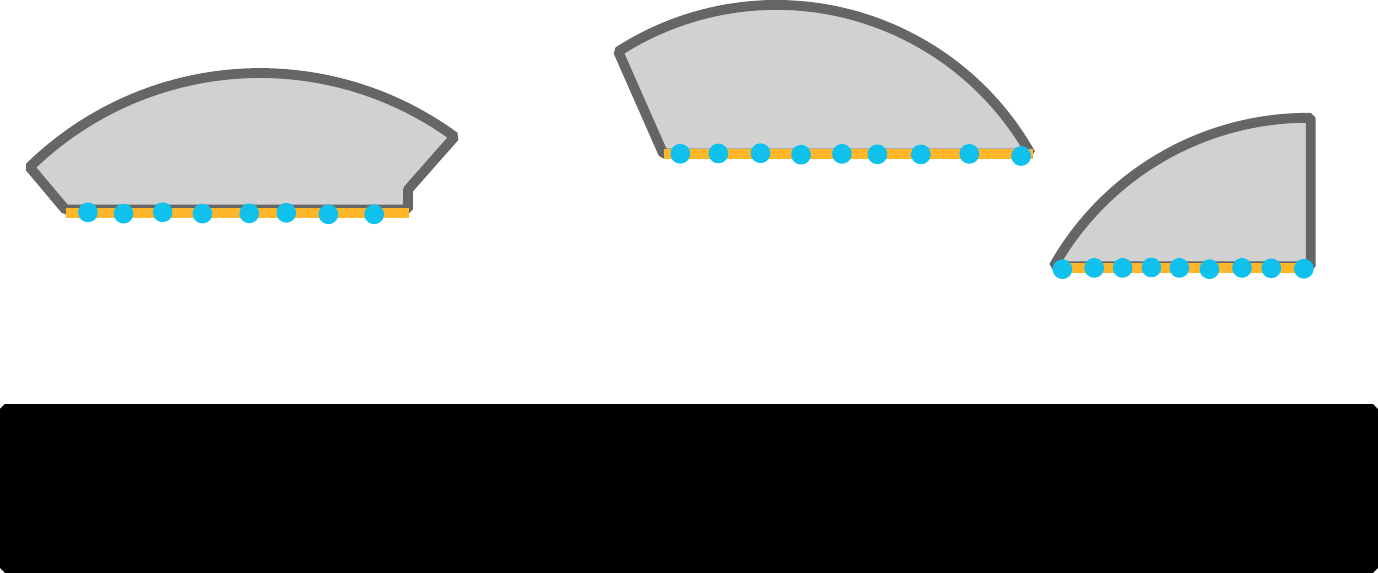}
      \subcaption{3D edge sampling}
    \end{subfigure}
    \hfill
    \begin{subfigure}{0.48\textwidth}
      \centering
      \includegraphics[width=0.6\textwidth]{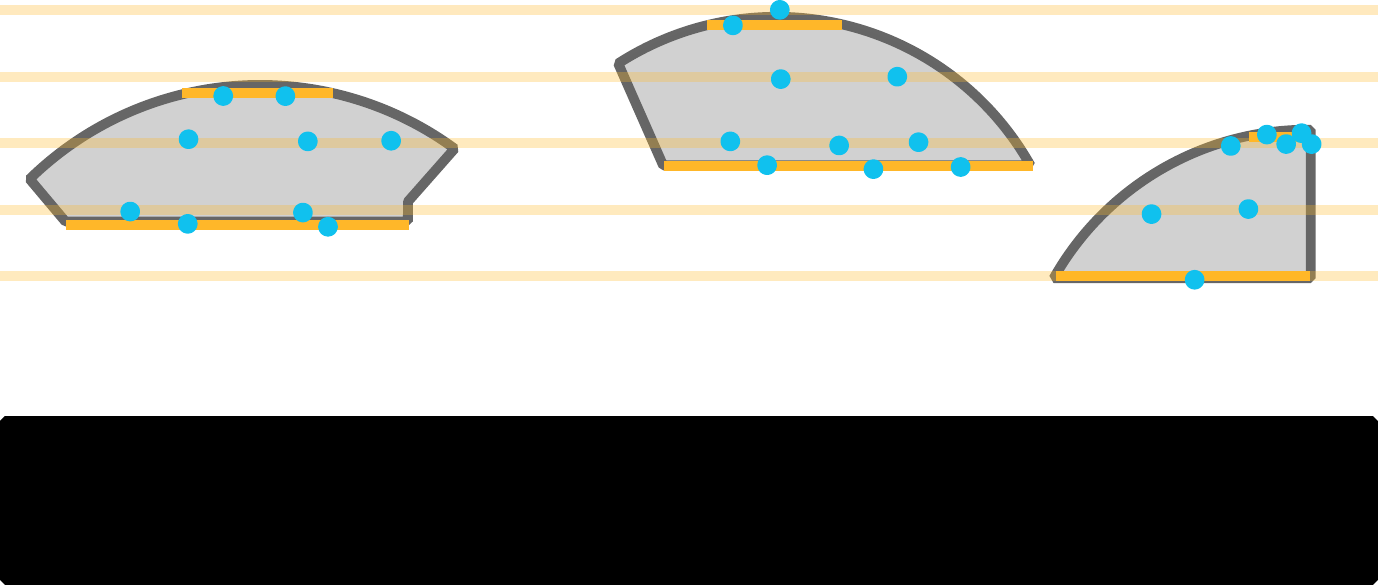}
      \subcaption{Global weighted face sampling}
    \end{subfigure}
    \caption{Visualizations of the four different sampling algorithms: \emph{optimized altitude with entry pose sampling}, \emph{random face sampling}, \emph{3D edge sampling}, and \emph{global weighted face sampling}. Two-dimensional representations of the visibility volumes are in gray, altitude slices are orange lines, and sampled configurations are  blue circular markers.}
  \label{fig:sampling}
\end{figure}

A more sophisticated approach developed by Obermeyer et al. \cite{Obermeyer2012} considers the fixed altitude slices of the target visibility volumes (i.e., planar visibility polygons). Vehicle configurations in each visibility polygon are sampled and a DTSPN \cite{Obermeyer2012, Isaacs.ACC.2011} is formulated to visit one configuration in each visibility polygon.
In \cite{Obermeyer2012}, two sampling algorithms were proposed: entry pose sampling---wherein samples are made along the edge of the polygon with heading angles that are tangent or inward pointing (Fig.~\ref{fig:dtsp}b)---and interior pose sampling---wherein samples are placed uniformly in a grid on the interior of the visibility polygons with uniformly sampled heading angles.
In \cite{Obermeyer2012}, entry pose sampling gave lower cost solutions than interior pose sampling. Thus, the entry pose sampling method is adopted here.
The constraints of the view planning problem \eqref{eq:constr1}--\eqref{eq:constr3} allow for visibility volumes to occupy disjoint segments of altitude. That is, there may not exist an altitude $z^* \in [z_{\rm min}, z_{\rm max}]$ that is common to all visibility volumes. While this does not pose an issue for some of the 3D algorithms proposed later, these cases cannot be solved by the 2D (constant-altitude) algorithms described here. However, introducing the additional constraint 
\begin{equation}
    h_{\rm max} + h_{\rm view} \leq d_{\rm max}
    \label{eq:extra_constraint_vis}
\end{equation}
ensures that the  visibility volumes for a target located on the ground plane and for a target located atop the highest object have at least one common altitude at $z^* = d_{\rm max}$. In general, there is a range of admissible altitudes $z^*$ that may be chosen. The choice of altitude impacts the 2D DTSPN algorithm since visibility polygons change in shape and size as the altitude varies. Intuitively, larger polygons are preferred over smaller ones since this increases the set of candidate configurations. This work proposes to identify an optimal working altitude for the 2D algorithm as follows. First, $n_{\rm slice}$ polygons are generated from each visibility volume mesh (i.e., for all targets) using the method described in \cite{Jiantao2004}.
Let $\mathcal{P} = \texttt{polygonFromMesh}(\hat{V},z)$ denote the polygon that results from slicing mesh $\hat{V}$ at altitude $z$ and let $\texttt{polygonArea}(\mathcal{P})$ denote the corresponding area.
The optimal altitude $z^*$ is chosen as the one that maximizes the sum of polygonal areas across all visibility polygons:
\begin{equation}
    z^*=\underset{z \in Z}{\rm argmax}
    \sum^{M-1}_{i=0} \texttt{polygonArea}(\texttt{polygonFromMesh}(\hat{V}_i,z))
\end{equation}
where $Z= \{z_0, \ldots, z_{n_{\rm{slice} - 1}} \}$ is a set of altitudes at which the polygons are computed.  Note that all altitudes $z \in Z$ are constrained such that 
$z_{\rm min}\leq z \leq z_{\rm max}$. Also, note that $z_0$ and $z_{\rm{slice} - 1}$ are not $z_{\rm min}$ and $z_{\rm max}$ respectively, instead $z_0$ and $z_{\rm{slice} - 1}$ are slightly offset into the body to avoid floating point error. The selection of $z^*$ is visualized in Fig~\ref{fig:sampling}a.



\section{3D Algorithms}
\label{sec:proposed}

Inspection tours that admit three-dimensional maneuvering can potentially lead to path length reductions when compared to two-dimensional (constant-altitude) tours.   
The solution techniques in this work all use a transformation approach to solve the (2D or 3D) DTSPN according to the following steps: compute the approximation of the target visibility volume (Sec.~\ref{sec:volumeApproximation}), sample the visibility volumes to create graph vertices corresponding to vehicle configurations, calculate edge costs between vertices using the 3D Dubins path planning algorithm, and solve for a DTSPN tour. 
Section~\ref{sec:proposed:sampling} details three algorithms to sample the target visibility volumes: random face sampling, 3D edge sampling, and global weighted face. 
Section~\ref{sec:angleBisector} then introduces a heuristic approach that improves the edge cost computation time using a modified Euclidean distance edge cost and a geometric approach to assign heading and pitch angles at each configuration in the tour.
\subsection{Sampling Strategies}
\label{sec:proposed:sampling}
\subsubsection{Random Face Sampling}
\label{sec:rface}
The \emph{random face sampling} algorithm extends the 2D entry pose strategy from \cite{Obermeyer2012} to sample 3D vehicle configurations across the surface of the target visibility volume with a uniform distribution. The approach is detailed in Algorithm~\ref{alg:faceSample} and visualized in Fig.~\ref{fig:sampling}b. The algorithm randomly finds $n_{\rm pts}$ three-dimensional points on the faces of each triangular mesh in the set of triangular meshes $\hat{V}_{0:M-1}=\{ \hat{V}_1, \ldots, \hat{V}_{M-1} \}$ and assigns to each point a set of configurations with  $n_\psi$ and $n_\gamma$ unique heading and pitch angles, respectively. The sampling method returns a total of  $n_{\rm pts}n_\psi n_\gamma$ vehicle configurations per visibility volume. 
First, the set of configurations $\mathcal{Q}$ is initialized as an empty set, and the area of each face in the mesh is calculated (lines \ref{alg:faceSample:face0}-\ref{alg:faceSample:face1}).
The area of each triangular face element, $a_{ij}$, is calculated by the \texttt{elementArea} function using
\begin{equation}
  a_{ij} = \texttt{elementArea}(\hat V_{ij}) =\frac{1}{2}||(\bm{c}_{ij}^0 - \bm{c}_{ij}^1) \times (\bm{c}_{ij}^0 - \bm{c}_{ij}^2)||\;,
\end{equation}
where $\times$ is the vector cross product and ${\bm c}_{ij}^0, {\bm c}_{ij}^1$, and ${\bm c}_{ij}^2 $ are the three vertices contained in the triangular face element $\hat{V}_{ij}$.
Next, the proportion of each face area to the total surface area of the mesh $\hat{V_i}$ is calculated using element-wise division (line~\ref{alg:faceSample:rand0}).
The \texttt{randomSetOfIndices} function identifies $n_{\rm pts}$ random faces by sampling faces with probability in proportion to the weights ${\bm w}_F$ (line~\ref{alg:faceSample:rand1}).
The use of the proportional surface area during the random selection process gives every point on the surface of the target visibility region an equal chance of being selected.
For each selected triangular face, a point on the face is randomly selected using a Barycentric coordinate system \cite{Turk1990} (lines~\ref{alg:faceSample:bary0}-\ref{alg:faceSample:bary1}).
The Barycentric coordinate system allows for the mapping of two random numbers $r_0$ and $r_1$ sampled uniformly from the interval $[0,1]$ onto a triangle, embedded in $\mathbb{R}^3$, with the weighted sum of its vertices \cite{Turk1990}.
The random numbers $r_0$ and $r_1$ are first sampled (line~\ref{alg:faceSample:uniformRand}) and then a position on the chosen triangular face is determined (line~\ref{alg:faceSample:bary1}).
For each position, $n_\psi$ heading angles sampled uniformly between $0$ and $2\pi$ as well as $n_\gamma$ pitch angles between $\gamma_{\rm min}$ and $\gamma_{\rm max}$ are sampled uniformly then added to the set of vehicle configurations. 
The runtime of the algorithm is dominated by the nested for loops on lines \ref{alg:faceSample:for0} and \ref{alg:faceSample:for2} running $M|\hat V|_{\rm max}$ times---where $|\hat{V}|_{\rm max} = {\rm max}_{i \in\{0,1,\ldots,M-1\}} (|\hat{V}_i|)$ is the maximum number of faces in a single mesh---and the collections of nested loops on lines \ref{alg:faceSample:heading0}-\ref{alg:faceSample:for1End} which run $Mn_{\rm pts} n_\psi n_\gamma$.
 When the number of mesh faces in a visibility volume is greater than the number of samples collected $ |\hat{V}|_{\rm max} > n_{\rm pts}n_\psi n_\gamma$ the runtime is $O(M|\hat V|_{\rm max})$.
\begin{algorithm}[h]
  \caption{Random Face Sampling}
  \begin{algorithmic}[1] 
    \STATEx \hspace{-2.1em}  {\bf function:} \texttt{RandomFaceSampling}$(\hat{V}_{0:M-1}, n_{\rm pts}, n_\psi, n_\gamma, \gamma_{\rm max}, \gamma_{\rm min})$
    \STATEx \hspace{-2.1em}   {\bf input:} target visibility volume mesh $\hat{V}_{0:M-1}$, number of points to sample $n_{\rm pts}$, number of heading angles $n_{\psi}$, number of pitch angles $n_\gamma$, max pitch angle $\gamma_{\rm max}$, min pitch angle $\gamma_{\rm min}$
    \STATEx \hspace{-2.1em}  {\bf output:} a set of vehicle configurations for each target $\mathcal{Q}$
    \STATE $\mathcal{Q} \gets \emptyset$
    \FOR{$\hat V_i \in \hat V_{0:M-1}$} \label{alg:faceSample:for0}
        \STATE $\mathcal{Q}_i \gets \emptyset, {\bm a}_i \gets \emptyset$ \label{alg:faceSample:face0}
        \FOR {$ \hat{V}_{ij} \in \hat{V_i}$} \label{alg:faceSample:for2}
          \STATE ${\bm a}_i \gets {\bm a}_i \cup \texttt{elementArea}(\hat{V}_{ij})$ \label{alg:faceSample:area}
        \ENDFOR \label{alg:faceSample:face1}
        \STATE ${\bm w}_F = {{\bm a}_i}/{\sum_{j=0}^{|{\bm a}_i| - 1}a_{ij}}$ \label{alg:faceSample:rand0}
        \STATE $I \gets \texttt{randomSetOfIndices}(n_{\rm pts}, {\bm w}_F )$ \label{alg:faceSample:rand1}
        \FOR {$i \in I$}
          \STATE $\bm{c}^{\rm 0}_{ij}, \bm{c}^{\rm 1}_{ij}, \bm{c}^{\rm 2}_{ij} \gets \texttt{getVertices}(\hat{V}_{ij})$ \label{alg:faceSample:bary0}
          \STATE $r_{\rm 0} \sim \mathcal{U}_{[0,1]}$, $r_{\rm 1} \sim \mathcal{U}_{[0,1]}$ 
          \label{alg:faceSample:uniformRand}
          \STATE $\bm{s} \gets \bm{c}^{\rm 0}_{ij}(1 - \sqrt{r_{\rm 0}}) + \bm{c}^{\rm 1}_{ij}\sqrt{r_{\rm 0}}(1 - r_{\rm 1}) + \bm{c}^{\rm 2}_{ij} \sqrt{r_{\rm 0}}r_{\rm 1}$ \label{alg:faceSample:bary1}
          \FOR {$j \in \rangeset{n_{\psi} - 1}$}\label{alg:faceSample:heading0}
            \FOR {$k \in  \rangeset{n_{\gamma} - 1}$} \label{alg:faceSample:for1End}
              \STATE $\mathcal{Q}_i \gets \mathcal{Q}_i \cup \left(\bm{s}, 2j\pi/n_\psi , \gamma_{\rm min} + k(\gamma_{\rm max} - \gamma_{\rm min})/\texttt{max}(n_\gamma - 1, 1)\right)$
            \ENDFOR
          \ENDFOR \label{alg:faceSample:heading1}
        \ENDFOR 
        \STATE $\mathcal{Q} \gets \mathcal{Q} \cup \mathcal{Q}_i$
    \ENDFOR
  \end{algorithmic}
\label{alg:faceSample}
\end{algorithm}
\subsubsection{3D Edge Sampling}
\label{sec:e3d}
The second sampling strategy proposed is \emph{3D edge sampling} wherein the 2D entry pose strategy from \cite{Obermeyer2012} is extended to sample 3D vehicle configurations across the lowest feasible altitude. For the visibility volume shapes studied here this is also the altitude where the cross-sectional area is largest for each shape.
The 3D edge sampling algorithm, detailed in Algorithm~\ref{alg:e3d} and visualized in Fig.~\ref{fig:sampling}c, finds $n_{\rm pts}$ three-dimensional points on the polygon created by slicing the triangular mesh along the lowest feasible altitude and distributing points uniformly along the perimeters. The algorithm then assigns a set of configurations to each point with  $n_\psi$ and $n_\gamma$ unique heading and pitch angles, respectively. The sampling method returns a total of  $n_{\rm pts}n_\psi n_\gamma$ vehicle configurations per visibility volume. 
First, the set of configurations $\mathcal{Q}$ is initialized as an empty set (line \ref{alg:e3d:init}).
Then, for each visibility volume a subset of points contained in that volume is initialized (line \ref{alg:e3d:init_i}).
Next, the $z$ minimum altitude for the triangular mesh is found by finding the minimum height coordinate in the set of vertices in the mesh $\hat{V}^z_i$(line \ref{alg:e3d:zbounds}).
After, the $\texttt{polygonFromMesh}$ algorithm
takes the triangular mesh and the $z_{\rm min}$ altitude and returns a polygonal slice of the mesh 
(line~\ref{alg:e3d:polygon}). A set of points, $\bm{\lambda} \in \mathbb{R}^2$, placed uniformly along the edge of the polygon is found using the \texttt{uniformPerimeterPoints} which takes a polygon and the number of points desired as arguments (line \ref{alg:e3d:uniformPerimeter}). Note that lines~\ref{alg:e3d:zbounds}--\ref{alg:e3d:uniformPerimeter} can be modified to produce samples at multiple altitude slices if desired.
Next, the algorithm iterates through each sampled point and assigns heading and pitch angles. To ensure inward-pointing heading angles, the direction of the line segment containing the sample point is found using the $\texttt{tangentAngle}$ function (line \ref{alg:e3d:heading}).
The points in the polygon defined by \texttt{polygonFromMesh} have a positive signed area. Thus, the inward-pointing heading angles are the angles from $[0, \pi]$ measured counter-clockwise from the tangent angle.
For each position, $n_\psi$ heading angles between $\psi_q$ and $\psi_q + \pi$ and $n_\gamma$ pitch angles between $\gamma_{\rm min}$ and $\gamma_{\rm max}$ are sampled uniformly and returned as part of the vehicle configurations (lines \ref{alg:e3d:configs0}-\ref{alg:e3d:configs1}). 
To achieve $n$ equally spaced angle samples, including the minimum and maximum angle, the range is divided into $n-1$ sub-sections. The $\texttt{max}$ function, on lines \ref{alg:e3d:headings1} and \ref{alg:e3d:pitch}, ensures the range is never divided by zero (the case where $n_\gamma$ or $n_\psi$ is one).
The runtime complexity is dominated by the for loops on lines 
\ref{alg:e3d:sample0}-\ref{alg:e3d:sample1} which have a worst-case running time complexity of $O(Mk)$, where $k=|\hat V|_{\rm max}^2 + n_{\rm pts}n_\psi n_\gamma$. $|\hat V|_{\rm max}^2$ is the runtime of the $\texttt{polygonFromMesh}$ algorithm while $n_{\rm pts}n_\psi n_\gamma$ is the runtimes for the nested for loops (lines \ref{alg:e3d:configs0}-\ref{alg:e3d:configs1}). For a typical choice of parameters, the number of faces in the target visibility volume mesh squared is greater than the total number of configurations returned, $|\hat V|_{\rm max}^2 > n_{\rm pts} n_\gamma n_\psi$ and the overall time-complexity is $O(M|\hat V|_{\rm max}^2)$.
\begin{algorithm}[h]
  \caption{3D Edge Sampling}
  \begin{algorithmic}[1] 
    \STATEx \hspace{-2.1em}  {\bf function:} \texttt{3DEdgeSampling}$(\hat{V}_{0:M-1}, n_{\rm pts}, n_\psi, n_\gamma, \gamma_{\rm max}, \gamma_{\rm min})$
    \STATEx \hspace{-2.1em}   {\bf input:} target visibility volume mesh $\hat{V}_{0:M-1}$, number of points to sample $n_{\rm pts}$, number of heading angles $n_{\psi}$, number of pitch angles $n_\gamma$, max pitch angle $\gamma_{\rm max}$, min pitch angle $\gamma_{\rm min}$
    \STATEx \hspace{-2.1em}  {\bf output:} a set of vehicle configurations for each target $\mathcal{Q}$
    \STATE $\mathcal{Q} \gets \emptyset$ \label{alg:e3d:init}
    \FOR {$\hat{V}_i \in \hat{V}_{0:M-1}$} \label{alg:e3d:sample0}
        \STATE $\mathcal{Q}_i \gets \emptyset$\label{alg:e3d:init_i}
        \STATE $z_{\rm min} \gets \texttt{min}(\hat{V}^z_i)$ \label{alg:e3d:zbounds}
        \STATE $\mathcal{P} \gets \texttt{polygonFromMesh}(z_{\rm min}, \hat{V}_i)$ \label{alg:e3d:polygon}
        \STATE $\{\bm{\lambda}_0, \ldots, {\bm \lambda}_{{n_{\rm pts}}-1}\} \gets \texttt{uniformPerimeterPoints}(\mathcal{P}, n_{\rm pts})$ \label{alg:e3d:uniformPerimeter}
        \FOR {$m \in \rangeset{n_{\rm pts} - 1}$} \label{alg:e3d:ni}
              \STATE $\psi_q \gets \texttt{tangentAngle}({\bm \lambda}_m, \mathcal{P})$ \label{alg:e3d:heading}     
              \FOR {$j \in \rangeset{n_\psi - 1}$} \label{alg:e3d:configs0}
               \STATE $\psi \gets \psi_q + {j\pi}/{\texttt{max}(n_\psi - 1, 1)}$ \label{alg:e3d:headings1}
                \FOR {$k \in \rangeset{n_\gamma - 1}$} \label{alg:e3d:headings0}
                   \STATE $\gamma \gets \gamma_{\rm min} + {k(\gamma_{\rm max}-\gamma_{\rm min})}/{\texttt{max}(n_\gamma - 1, 1)}$ \label{alg:e3d:pitch}
                   \STATE $\mathcal{Q}_i \gets \mathcal{Q}_i \cup \left(\bm{\lambda}_m,\: z_{\rm min}, \: \psi, \: \gamma\right)$
                \ENDFOR
              \ENDFOR \label{alg:e3d:configs1}
        \ENDFOR
      \STATE $\mathcal{Q} \gets \mathcal{Q}_i \cup \mathcal{Q}$
    \ENDFOR\label{alg:e3d:sample1}
  \end{algorithmic}
\label{alg:e3d}
\end{algorithm}
\subsubsection{Global Weighted Face Sampling}
\label{sec:globalFace}
The third proposed sampling strategy is \emph{global weighted face sampling}.
Rather than sampling the visibility volumes at the lowest altitude, all target visibility volumes are sampled along a common set of altitude planes and the number of samples allocated to each plane is determined by the cross-sectional perimeter distribution of each altitude summed across all target visibility volumes. This approach places more samples at altitudes common to all targets that, on average, also have large cross-sectional areas. 
This sampling method is detailed in Algorithm~\ref{alg:globalFace} and visualized in Fig.~\ref{fig:sampling}d.
The algorithm takes a set of target visibility meshes $\hat V_{0:M-1}$ and returns a set of vehicle configurations $\mathcal{Q}$ for each mesh given the parameters $n_{\rm pts}$, $n_\psi$, $n_\gamma$, $n_{\rm slice}$, $\gamma_{\rm max}$, and $\gamma_{\rm min}$ where $n_{\rm slice} \geq 2$ is the number of altitude slices to consider.
Let $\hat{V}_{0:M-1}^z$ denote the set of all $z$ heights for every vertex contained across the $M$ meshes $\hat V_{0:M-1}$. 
First, the global minimum, the global maximum altitude, and the slicing altitude step size are found (lines \ref{alg:globalFace:bounds}-\ref{alg:globalFace:step}).
Then a vector $\bm{\mu}$ is initialized with zeros, denoted as $0_{n_{\rm slice}\times 1}$ (line~\ref{alg:globalFace:init}), and later stores the total perimeter summed across all visibility polygons at the corresponding altitude slice. The target visibility volumes are sliced into polygons with fixed altitude (i.e. parallel to the $xy$ plane) using the \texttt{polygonFromMesh} function, lines \ref{alg:globalFace:slice0}-\ref{alg:globalFace:slice1}.
The lowest $z$ plane is the visibility volumes' global minimum $z$ height ($\zeta_{\rm min}$) and the highest $z$ plane is the visibility volumes' global maximum $z$ height ($\zeta_{\rm max}$), line \ref{alg:globalFace:bounds}. The nominal set of altitude planes is then $Z = \{ z_0, \ldots, z_{n_{\rm slice}-1} \}$ where $z_0=\zeta_{\rm min}$, $z_{n_{\rm slice}-1} =\zeta_{\rm max}$ and $z_{i+1} - z_{i} = L$.
At each plane $z\in Z$, polygons are created from the target visibility volume and the polygons' perimeters are accumulated, line \ref{alg:globalFace:accumArea}.
The sample points in each $z$ plane are then distributed in proportion to the accumulated perimeters, lines \ref{alg:globalFace:sample0}-\ref{alg:globalFace:sample1}.
The function \texttt{iteratePerimeters} takes six arguments: the mesh to iterate across, a perimeter distribution, the minimum altitude, the maximum altitude, the step size, and the total number of sample points. It returns a variable number of $n_z \leq n_{\rm slice}$ elements where each element is a pair consisting of a $z_r$ altitude and the number of points to sample at that altitude, $n_r$.
An altitude slice $z_r$  is either an element of $Z$ and/or an altitude located at the top or bottom of each visibility volume.
At each altitude $z_r$ the corresponding value of  ${\bm \mu}$ is determined (or interpolated, in the special case that $z_r \notin Z$) and the $n_{\rm pts}$ are distributed to each $z_r$ in proportion to the result.
In the event that no slices intersect the visibility mesh then $n_z = 2$ and the heights $z_r$ are returned corresponding to the top and bottom of the target visibility volume.
Next, samples $\bm{\lambda} \in \{\bm{\lambda}_0, ..., \bm{\lambda}_{n_r -1}\}$ are placed uniformly around the perimeter of each polygon created by the intersection of the $z_r$ planes and the target visibility volume with the function $\texttt{uniformPerimeterPoints}$, line \ref{alg:globalFace:uniformPerimeter}.
The heading and pitch angles are sampled in the same way as entry pose sampling \cite{Obermeyer2012}, pointing tangent or inward with respect to the polygon.
The angle tangent to each point $\bm{\lambda}$ on the perimeter of the polygon is found with the $\texttt{tangentAngle}$ function.
The pitch angles are uniformly sampled within the pitch angle constraints.
The runtime complexity is dominated by the for loops on lines 
\ref{alg:globalFace:sample0}-\ref{alg:globalFace:sample1} which have a worst-case running time complexity of $O(Mn_{\rm slice}k)$, where $k=|\hat V|_{\rm max}^2 + n_rn_\gamma n_\psi$. For a typical choice of parameters, the number of faces in the target visibility volume mesh squared is greater than the total number of configurations returned, $|\hat V|_{\rm max}^2 > n_r n_\gamma n_\psi$ and the overall time-complexity is $O(Mn_{\rm slice}|\hat V|_{\rm max}^2)$.
\begin{algorithm}[h!]
  \caption{Global Weighted Face}
  \begin{algorithmic}[1]
    \STATEx \hspace{-2.1em}  {\bf function:} \texttt{GlobalWeightedFace}$(\hat {V}_{0:M-1}, n_{\rm pts}, n_\psi, n_\gamma, n_{\rm slice}, \gamma_{\rm max}, \gamma_{\rm min})$
    \STATEx \hspace{-2.1em}   {\bf input:} set of triangular meshes $\hat{V}_{0:M-1}$, number of points to sample $n_{\rm pts}$, number of heading angles $n_{\psi}$, number of pitch angles $n_\gamma$, number of altitude slices $n_{\rm slice}$, max pitch angle $\gamma_{\rm max}$, min pitch angle $\gamma_{\rm min}$
    \STATEx \hspace{-2.1em}  {\bf output:} a set of vehicle configurations for each target $\mathcal{Q}$
    \STATE $\zeta_{\rm max} \gets \texttt{max}(\hat {V}_{0:M-1}^z),\; \zeta_{\rm min} \gets \texttt{min}(\hat {V}_{0:M-1}^z)$ \label{alg:globalFace:bounds}
    \STATE $L \gets (\zeta_{\rm max} - \zeta_{\rm min})/(n_{\rm slice} - 1)$ \label{alg:globalFace:step}
    \STATE $\bm{\mu} \gets 0_{n_{\rm slice}\times 1}$ \label{alg:globalFace:init}
    \FOR {$i \in \rangeset{n_{\rm slice} - 1}$} \label{alg:globalFace:slice0}
      \FOR {$\hat{V}_i \in \hat{V}_{0:M-1}$}
        \STATE $\mathcal{P} \gets \texttt{polygonFromMesh}(\zeta_{\rm min} + Li, \hat {V}_i)$ \label{alg:globalFace:makePolys}
        \STATE $\mu_{i} \gets \mu_{i} + \texttt{perimeter}(\mathcal{P})$ \label{alg:globalFace:accumArea}
      \ENDFOR
    \ENDFOR  \label{alg:globalFace:slice1}
    \STATE $\mathcal{Q} \gets \emptyset$
    \FOR {$\hat{V}_i \in \hat{V}_{1:M}$} \label{alg:globalFace:sample0}
    \STATE $(z_r, n_r)_{r=0}^{n_{z}-1} \gets \texttt{iteratePerimeters}(\hat{V}_i, {\bm \mu}, \zeta_{\rm min}, \zeta_{\rm max}, L, n_{\rm pts})$
    \STATE $\mathcal{Q}_i \gets \emptyset$
    \label{alg:globalFace:iteratePerimeters}
      \FOR {$r \in \rangeset{n_{z}-1}$} 
        \STATE $\mathcal{P} \gets \texttt{polygonFromMesh}(z_r, \hat{V}_i)$
        \STATE $\{\bm{\lambda}_0, \ldots, {\bm \lambda}_{{n_r}-1}\} \gets \texttt{uniformPerimeterPoints}(\mathcal{P}, n_r)$ \label{alg:globalFace:uniformPerimeter}
        \FOR {$m \in \rangeset{n_r - 1}$} \label{alg:globalFace:ni}
              \STATE $\psi_q \gets \texttt{tangentAngle}({\bm \lambda}_m, \hat{V}_i)$        
          \FOR {$j \in \rangeset{n_\psi - 1}$}
            \FOR {$k \in \rangeset{n_\gamma - 1}$} \label{alg:globalFace:headings0}
              \STATE $\psi \gets \psi_q + {k\pi}/{\texttt{max}(n_\psi - 1, 1)}$ \label{alg:globalFace:headings1}
              \STATE $\gamma \gets \gamma_{\rm min} + {k(\gamma_{\rm max}-\gamma_{\rm min})}/{\texttt{max}(n_\gamma - 1, 1)}$
              \STATE $\mathcal{Q}_i \gets \mathcal{Q}_i \cup \left(\bm{\lambda}_m,\: z_r, \: \psi, \: \gamma\right)$
            \ENDFOR
          \ENDFOR
        \ENDFOR
      \ENDFOR
      \STATE $\mathcal{Q} \gets \mathcal{Q}_i \cup \mathcal{Q}$
    \ENDFOR \label{alg:globalFace:sample1}
  \end{algorithmic}
  \label{alg:globalFace}
\end{algorithm}
\subsection{Proposed Heuristics}
\subsubsection{Modified Euclidean Distance Traveling Salesperson Problem with Neighborhoods (METSPN)}
\label{sec:METSPN}
\begin{figure}[ht]
    \centering
    \includegraphics[width = 0.48\textwidth]{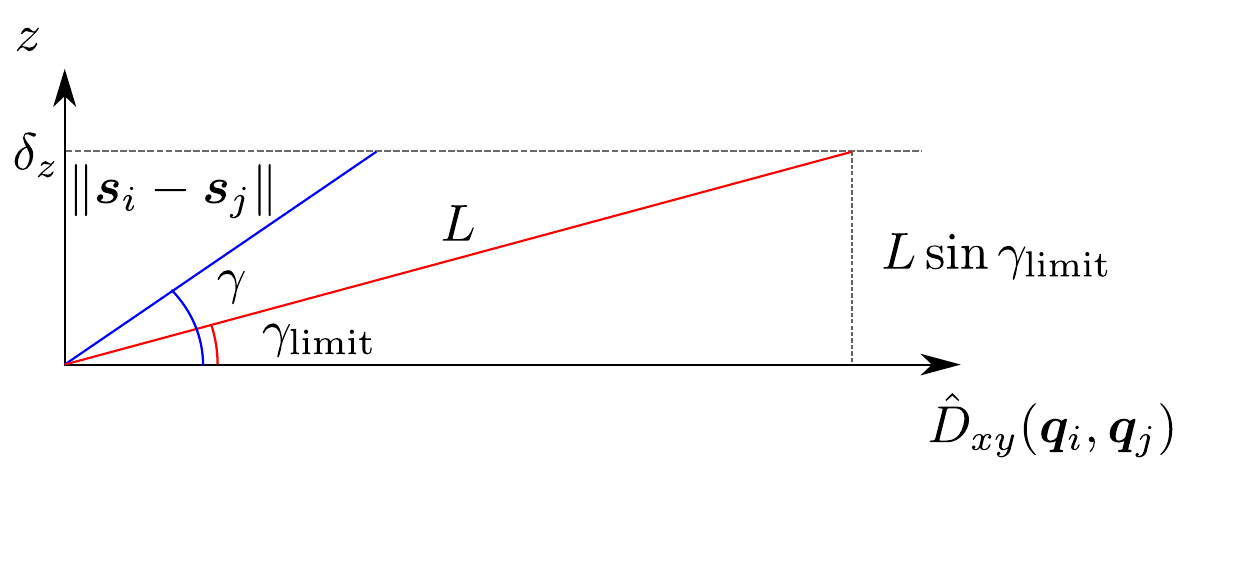}
    \caption{
        Visualization of the modified Euclidean distance. The Euclidean distance shown in blue with a pitch angle $\gamma > \gamma_{\rm limit}$ is  modified by extending the distance traveled in the $xy$ plane resulting in the red line with pitch angle $\gamma_{\rm limit}$. $\hat D_{xy}$ refers to the length of the Dubins path projected onto the $xy$ plane.
    }
    \label{fig:modifiedDistance}
\end{figure}

A bottleneck in the 3D DTSPN algorithms is the computation of the edge costs that require solving for a 3D Dubins path between two configurations ${\bm q}_i = (x_i, y_i, z_i, \psi_i, \gamma_i)$ and ${\bm q}_j = (x_j, y_j, z_j, \psi_j, \gamma_j)$.
Since the Dubins path is asymmetric the corresponding edge cost must be computed for each direction.
Here, we propose an approximation to this edge cost 
\begin{equation}
\hat D({\bm q}_i, {\bm q}_j) = {\rm max}\left(\frac{|\delta_z|}{\sin{\gamma_{\rm limit}}}, \lVert\bm{s}_i - \bm{s}_j\rVert_2\right)\;,
\label{eq:modified_dist}
\end{equation}
where $\delta_z = z_j - z_i$, $\gamma_{\rm limit} = \gamma_{\rm max}$ if $\delta_z > 0$ and $\gamma_{\rm limit}=\gamma_{\rm min}$ otherwise, $\bm{s}_i=(x_i, y_i, z_i)$ and $\bm{s}_j=(x_j, y_j, z_j)$.
The calculation is visualized in Fig.~\ref{fig:modifiedDistance}.
The distance \eqref{eq:modified_dist} is a lower bound on the actual 3D Dubins path length, i.e., $\hat D({\bm q}_0, {\bm q}_1) \leq D({\bm q}_0, {\bm q}_1)$, and is significantly faster to compute than solving for the Dubins path. Using this edge cost leads to a variant of the DTSPN we refer to as the modified Euclidean distance traveling salesperson problem (METSPN). Solving the METSPN gives a tour of 3D locations to visit. Once a tour is found for the METSPN it is converted into a feasible sequence of Dubins paths by assigning heading and pitch angles as follows. 
\subsubsection{Bisecting Angle Approximation}
\label{sec:angleBisector}
To assign heading and pitch angles a heuristic is adopted that extends the mean angle algorithm developed in  \cite{MacharetEtAl.DG.2012} to three dimensions. The approach is summarized in Algorithm~\ref{alg:angleBisector}. The proposed bisecting angle approximation takes as parameters: $\bm{V}$ a $M \times 3$ matrix corresponding to the sequence of vertices in the METSPN tour and the problem parameters: $\rho_{\rm min}$, $\gamma_{\rm min}$, and $\gamma_{\rm max}$.
The algorithm returns a set of vehicle configurations $\mathcal{Q}$ at each point in $\bm{V}$ with heading and pitch angles defined as the angle bisector of each consecutive triplet of vertices (for points spaced far apart) or as a straight segment (for points spaced close together).

\begin{figure}[ht]
    \centering
    \includegraphics[scale=.5]{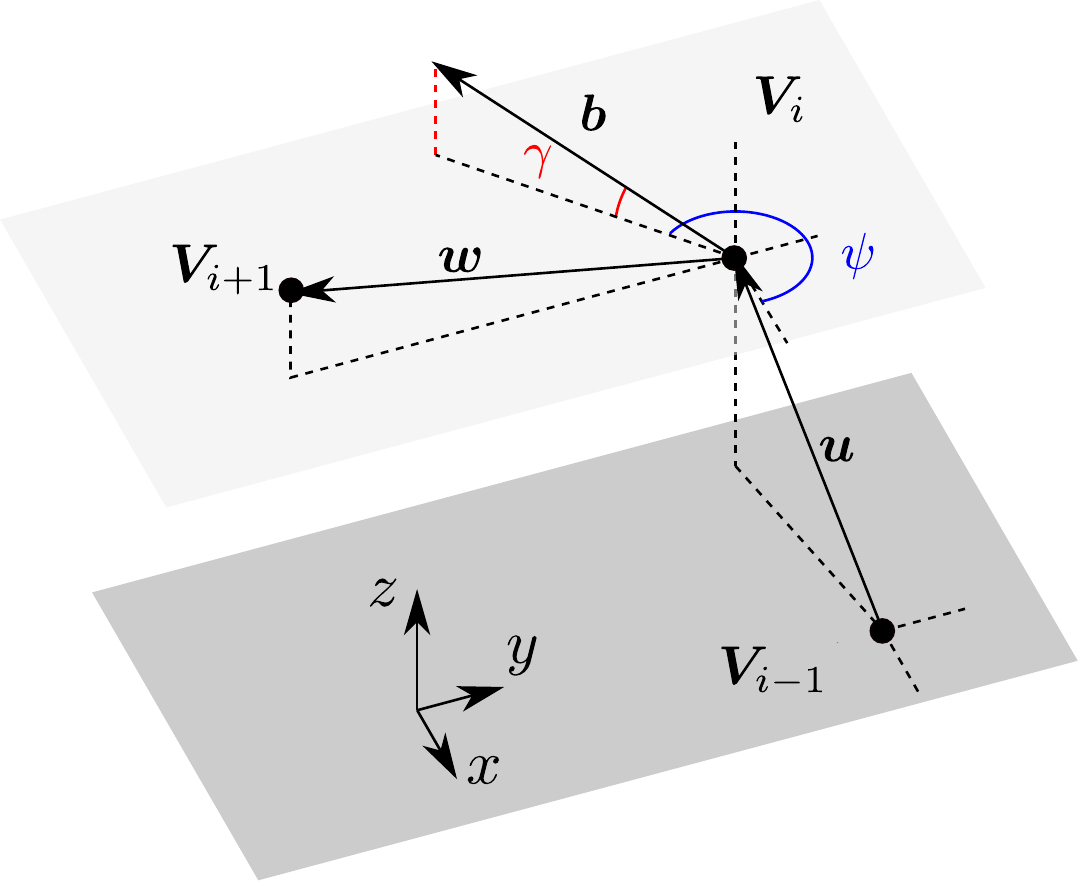}
    \caption{
    The notation used to determine the bisector vector for a triplet of three points: $\bm{V}_{i-1},{\bm V}_i, {\bm V}_{i+1}$. The orientation of the vectors  ${\bm u} = {\bm V}_i - {\bm V}_{i-1}$ and ${\bm w} = {\bm V}_{i+1} - {\bm V}_{i}$ are summed and normalized resulting in the vector $\bm{v}$. The heading angle $\psi$ is the component of $\bm{b}$ in the $xy$ plane while the pitch angle $\gamma$ is measured from the $xy$ plane.
    }
    \label{fig:bisector}
\end{figure}

To obtain the angle bisector at each vertex, calculate vectors from the preceding vertex $\bm{u} = {\bm V}_i - {\bm V}_{i-1} = ({u}_x, {u}_y, {u}_z)$ and to the following vertex $\bm{w}= {\bm V}_{i+1} - {\bm V}_{i}  = ({w}_x, {w}_y, {w}_z)$ (line \ref{alg:angleBisector:vector}). The vector $\bm{b} = {\bm w} + {\bm u} = ({b}_x, {b}_y, {b}_z)$ determines the heading angle $\psi$ in the $xy$ plane computed with the four-quadrant arctangent function (line \ref{alg:angleBisector:hbisector}).
A visualization of the calculation can be seen in Fig.~\ref{fig:bisector}.
The circular indexing of $\bm{V}$, a $M$ by 3 matrix, allows for the index $-1$ to refer to the last column of $\bm{V}$ and the index $n$ to refer to the first element of $\bm{V}$.
The pitch angle bisector is the angle between the vector $\bm{b}$ and the $xy$ plane (line~\ref{alg:angleBisector:fbisector}).
The resulting angle is saturated to be within the pitch angle bounds on line~\ref{alg:angleBisector:clip0}.
If vertices are close together then curve-curve-curve (CCC) Dubins paths may be created.
This should be avoided because the cost of (CCC) Dubins paths is much greater than the Euclidean distance.
The likelihood of CCC paths occurring is reduced by setting the heading and pitch in the direction of the line between two vertices.
If the distance between two vertices is small (less than the long path case in \cite{Shkel2001}), then heading and pitch angles are aligned with the while loop on lines \ref{alg:angleBisector:while0}-\ref{alg:angleBisector:while1}.
To align the headings of two configurations, the vector between the internal coordinates is found. The angle of this vector, $\bm{w}$, about the $z$ axis is used as the heading angle.
Then, the angle between the $xy$ plane and the vector $\bm{w}$ is found and saturated between $\gamma_{\rm min}$ and  $\gamma_{\rm max}$ to set the pitch angle.
Inside the loop, the index is advanced once but it is also advanced a second time if the current vertex and the next vertex are within $4\rho_{\rm min}$ units of each other (worst case for the long path case \cite{Shkel2001}).
The second index advance is required to pass over the next configuration because it was just modified.
\begin{algorithm}[ht]
  \caption{Bisect Angle Approximation}
  \begin{algorithmic}[1]
    \STATEx \hspace{-2.1em}  {\bf function:} \texttt{BisectAngleApprox}$(\bm{V}, \rho_{\rm min}, \gamma_{\rm min}, \gamma_{\rm max}$)
    \STATEx \hspace{-2.1em}   {\bf input:} $\bm{V}$ is a $n$ by 3 matrix of vertices that solve the METSPN, minimum turn radius $\rho_{\rm min}$, minimum pitch angle $\gamma_{\rm min}$, maximum pitch angle $\gamma_{\rm max}$
    \STATEx \hspace{-2.1em}  {\bf output:} set of configurations solving a DTSP $\mathcal{Q}$
    \STATE $\mathcal{Q} \gets \emptyset$
    \FOR {$i \in \{0, 1, 2 \dots M - 1\}$}
      \STATE $\bm{b} \gets \bm{V}_{i + 1} + \bm{V}_{i-1}$\text{// indexing into $\bm{V}$ is circular} \label{alg:angleBisector:vector}
      \STATE $\psi \gets \atantwo{{b}_x}{{b}_y}$ \label{alg:angleBisector:hbisector}
      \STATE $\gamma \gets \atantwo{{b}_z}{\sqrt{b_x^2 + b_y^2}}$ \label{alg:angleBisector:fbisector}
      \STATE $\gamma \gets\texttt{max}(\texttt{min}(\gamma,\gamma_{\rm max}), \gamma_{\rm min})$ \label{alg:angleBisector:clip0}
      \STATE $\mathcal{Q} \gets \mathcal{Q} \cup (\bm{V}_i,\psi,\gamma)$
    \ENDFOR
    \STATE $i \gets 0$
    \WHILE{$i < |\bm{V}|$} \label{alg:angleBisector:while0}
        \IF {$||\bm{V}_i - \bm{V}_{i+1}|| < 4\rho_{\rm min}$} \label{alg:angleBisector:close}
          \STATE $\bm{w} \gets \bm{V}_{i+1} - \bm{V}_{i}$
          \STATE $\psi \gets \atantwo{u_x,u_y}$
          \STATE $\gamma \gets \atantwo{{u}_z}{\sqrt{u_x^2 + u_y^2}}$
          \STATE $\gamma \gets\texttt{max}(\texttt{min}(\gamma,\gamma_{\rm max}), \gamma_{\rm min})$ \label{alg:angleBisector:clip1}
          \STATE $\mathcal{Q}_{i\psi} \gets \psi,~\mathcal{Q}_{(i+1)\psi} \gets \psi$
          \STATE $\mathcal{Q}_{i\gamma} \gets \gamma,~\mathcal{Q}_{(i + 1)\gamma} \gets \gamma$
          \STATE $i \gets i + 1$
        \ENDIF 
        \STATE $i \gets i + 1$
    \ENDWHILE \label{alg:angleBisector:while1}
  \end{algorithmic}
  \label{alg:angleBisector}
\end{algorithm}
\subsection{Illustrative Examples}
\begin{figure}[h]
    \begin{subfigure}{0.48\textwidth}
      \centering
      \includegraphics[width=\textwidth]{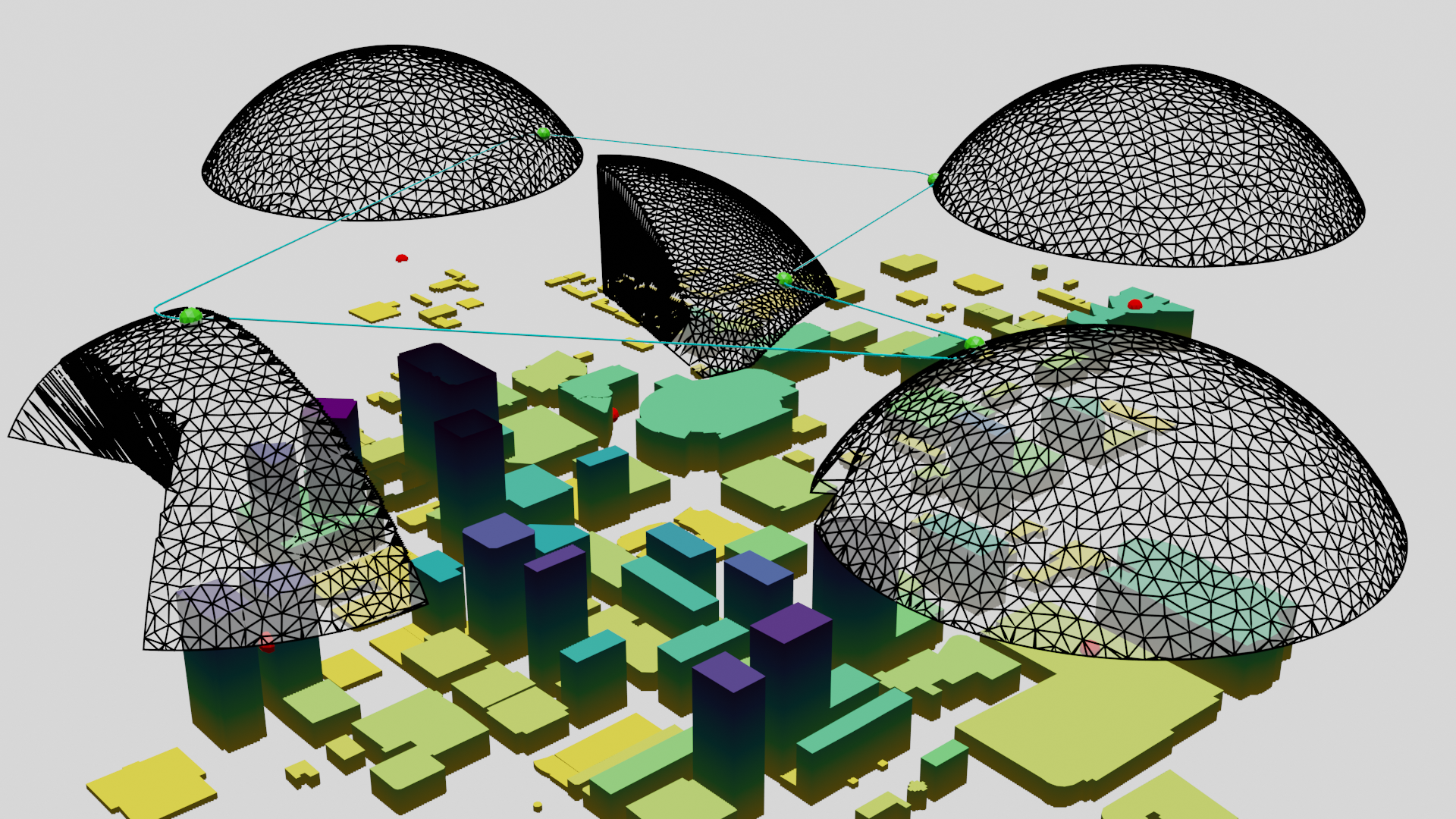}
      \subcaption{3D DTSP with neighborhoods (DTSPN) in Charlotte, North Carolina}
    \end{subfigure}
    \hfill
    \begin{subfigure}{0.48\textwidth} \label{fig:nyc}
      \centering
      \includegraphics[width=\textwidth]{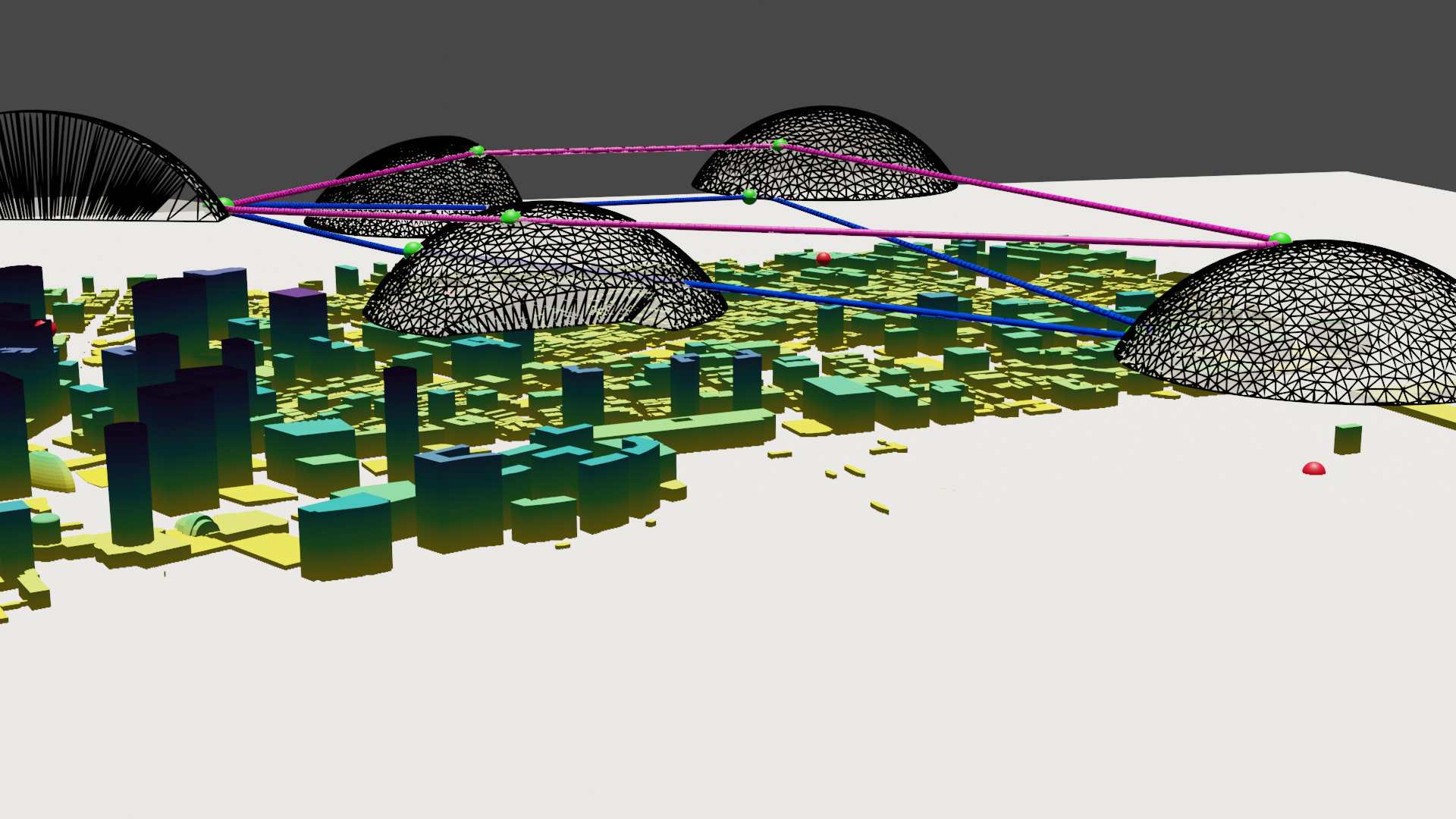}
      \subcaption{3D DTSP with neighborhoods (DTSPN) in New York City, New York}
    \end{subfigure}
  \centering
  \caption{
    Solutions to the 3D Dubins traveling salesperson problem with neighborhoods. Panel (a) was computed using the random face algorithm in light blue with 8 samples per target visibility volume, four heading angles per sample, and one pitch angle per sample-heading angle pair. Panel (b) was computed using the random face sampling algorithm in dark blue with $n_{\rm pts} = 32$ samples per target visibility volume, $n_\psi = 8$ heading angles per sample, and $n_\gamma = 3$ pitch angles per sample-heading pair; the two-dimensional entry pose sampling from \cite{Obermeyer2012} in magenta with $n_{\rm pts} = 32$ samples per target visibility volume and $n_\psi = 4$ heading angles per sample. The target visibility volume is translucent white with black edges and the targets are red spheres. The green spheres are the vehicle configurations for the solution to the DTSPN. The environment shown is a section of  New York City, New York obtained from the OpenStreetMap database. Building heights are indicated by the varying color scale from yellow to purple.
  }
  \label{fig:solutionsamples}
\end{figure}

An example of a view planning solution for five targets scattered around a city model of Charlotte, North Carolina is shown in Fig.~\ref{fig:solutionsamples}a. The example was constructed assuming a Dubins airplane model having a curvature radius of $\rho_{\rm min} = 40$ m  and pitch angle constraints $\gamma \in [-\pi/12, \pi/9]$. The random-face algorithm was used with $n_{\rm pts}=8$ samples per visibility volume, $n_{\psi} = 4$ heading angles per sample, and $n_{\gamma} = 1$ pitch angle per sample-heading angle pair.  The visibility volumes for targets that had no occlusions had a common dome shape, whereas targets located closer to objects had more arbitrary shapes. Another example Fig.~\ref{fig:solutionsamples}b illustrates the solution for five targets in a model of New York City, New York. This example compares the three-dimensional random-face algorithm with $n_{\rm pts} = 32$, $n_\psi = 8$, and $n_\gamma = 3$ pitch angles, to the two-dimensional optimized altitude entry pose sampling algorithm with $n_{\rm pts} = 32$, $n_\psi = 8$. The 3D path can change altitude which allowed the algorithm to find a lower cost path of 3920m while the 2D algorithm maintained constant altitude and found a path of cost 4285m, a 10.9\% reduction in path cost.

\section{Numerical Performance Study}
\label{sec:results}
The 2D algorithms from Sec.~\ref{sec:baseline} were compared to the 3D algorithms from Sec.~\ref{sec:proposed} through a Monte-Carlo experiment that randomized target locations and a number of targets located in an urban environment. This section describes the implementation of the algorithms, the design of the Monte-Carlo study, and discusses the results.
\subsection{Implementation}
The algorithms in this work were written in python 3.9 \footnote{The implementation of this study can be found at \url{https://github.com/robotics-uncc/VisualTour3DDubins}.}\cite{Python} using a number of packages, including Shapely \cite{shapely} for polygonal operations and NumPy \cite{2020NumPy-Array} for working with matrices.
The GLKH traveling salesperson solver \cite{GLKH} was used to solve the generalized traveling salesperson problems that arise from DTSPs.
The target visibility volumes were created with data from OpenStreetMap \cite{OpenStreetMap}, inverse depth calculations from the target location using OpenGL \cite{OpenGL}, and Blender \cite{blender} was used for intersecting the triangular meshes within the feasible airspace $F$ as well as decimating the meshes (i.e., reducing the number of triangular faces).
This work uses \cite{TangEtAl.SST.1998} to compute 2D Dubins paths for the 2D algorithms. 
The algorithm simulations were performed on an AMD Threadripper 3990X running Ubuntu 20.04 with one thread allocated to the algorithm.
\subsection{Monte-Carlo Experiment}
 A Monte-Carlo experiment was designed using the environments described in  Table~\ref{tbl:environment}. The environments were created by capturing all of the buildings in a rectangular area in New York City with the OpenStreetMap database and limiting the building heights to 300 m.
 \begin{table}[h!]
\begin{center}
  \caption{Description of environments obtained from an OpenStreetMap database for New York City, USA, and used for the Monte-Carlo experiment.}
   \label{tbl:environment}
\begin{threeparttable}
    \begin{tabular}{cccc}
        Number of Targets & Number of objects & Width & Depth\\  [0.5ex]
      \hline
      5 & 5624 & 1986 m & 2090 m\\      
      10 & 9202 & 2809 m & 2857 m\\ 
      15 & 11584 & 3440 m & 3621 m\\      
      20 & 12119 & 3972 m & 4181 m
    \end{tabular}
\end{threeparttable}
\end{center}
\end{table}
Target locations were randomized for each trial and determined by sampling the environment and placing targets on the ground, the wall of buildings, or the roofs of buildings according to a user-defined distribution.
The radius of the target visibility volumes was 300 m with each target being at least 600 m apart.
The proposed sampling methods and heuristics are independent and studied here in different combinations.
The algorithms parameters were varied as follows:
the number of samples per visibility volume was varied between $n_{\rm pts} = \{2, 4, 8, 16, 32\}$, the number of heading angles per sample was $n_\psi = \{2, 4, 8\}$.
To reduce the number of trials, only one pitch angle $(n_\gamma = 1)$ of  $0^\circ$ was used by passing $0^\circ$ for $\gamma_{\rm min}$ and $\gamma_{\rm max}$ to the random face sampling (Sec.~\ref{sec:rface}), 3D edge sampling (Sec.~\ref{sec:e3d}), and global weighted face sampling (Sec.~\ref{sec:globalFace}) algorithms.
The Dubins airplane had a minimum curvature radius of $\rho_{\rm min} = 40$ m and a pitch angle constrained between -${\pi}/{12}$ and ${\pi}/{9}$, similar to \cite{Vana2020}.
A total of 80 configurations of targets were generated, divided evenly among groups of 5, 10, 15, and 20 targets.
Every combination of algorithm parameters was evaluated with the 80 configurations. The normalized tour cost (total length of the tour divided by the turn radius) and the computation time were recorded. 
The algorithms are denoted by acronyms wherein the prefix is either 2D-DTSP, 2D-DTSPN, 3D-DTSPN, or 3D-METSPN corresponding to the algorithms of Sections \ref{sec:baselineDTSP}, \ref{sec:baselineDTSPN}, \ref{sec:proposed:sampling}, and \ref{sec:METSPN}, respectively.
The 2D-DTSP is followed by a dash and an integer representing the number of heading angles.
The remaining two algorithms are described by a sampling method acronym: entry pose sampling (ETRY) from Sec.~\ref{sec:baselineDTSPN}, random face sampling (RFAC) from Sec.~\ref{sec:rface}, 3D edge sampling (E3D) from Sec.~\ref{sec:e3d}, or global weighted face sampling (GWF) from Sec.~\ref{sec:globalFace} followed by a dash and an integer representing the number of heading angles and another dash and an integer representing the number of samples per target visibility volume (i.e., 2D-DTSPN-ETRY-4-16 corresponds to a 2D DSTPN using entry pose sampling with 4 heading angles and 16 sample points per target visibility region).
\begin{figure}[t]
  \centering
  \includegraphics[width = 0.75\textwidth]{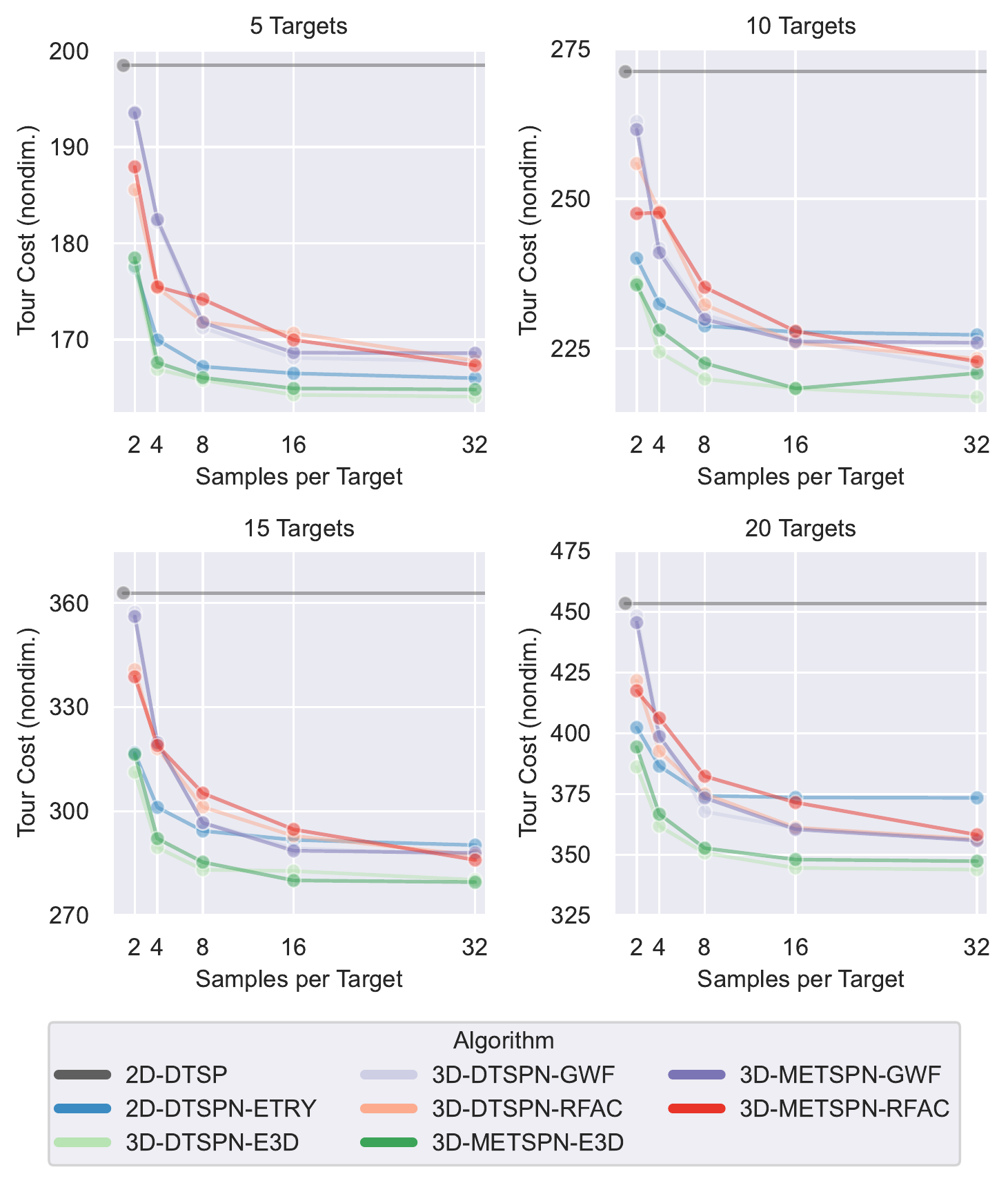}
  \caption{
    The line plots show the median non-dimensional tour cost of the different algorithms as the number of samples per target visibility volume increases.
    }
  \label{fig:costSelectedAlgorithms}
\end{figure}
\subsubsection{Analysis of Monte-Carlo Study}
In general, two-dimensional methods at a fixed altitude performed better if the targets are all located at similar heights; whereas, 3D methods trended towards better tour cost when targets occupy a wide range of altitudes. 
The median path length, normalized by dividing the cost by the minimum curvature radius, of each view-planning tour (i.e., cost) of the Monte-Carlo runs for an increasing number of targets, the number of heading angles is held at $n_\psi = 8$, and the number of pitch angles is held at $n_\gamma = 1$ is plotted in Fig.~\ref{fig:costSelectedAlgorithms}.
The DTSP algorithms that only visit a single point (gray) have one location sample per visibility volume but the lines were extended along the abscissa for comparison.
The DTSP is inefficient in our problem because shorter paths can be obtained between targets by flying through the boundary of their corresponding visibility volumes rather than requiring the paths to pass through the visibility volume centers.
As the number of heading angles increases the mean cost of the solution decreases, as expected.
The METSPN algorithms have a similar cost to the eight sampled heading angle solutions.
Most of the medians for different algorithms approach an asymptote, suggesting that they are converging towards a fixed median tour cost (i.e., further increasing the number of samples has diminishing returns).
For a large number of samples, the proposed random face sampling algorithm yields a lower tour cost than the optimized altitude 2D algorithm. However, the median of the 3D edge sampling algorithm is less than the optimal altitude 2D algorithm for all numbers of samples greater than 2. This may be due to the 3D algorithms spreading their samples across another dimension (altitude). The 3D algorithms that spread the samples along the vertical dimension of each visibility volume perform worse than the algorithm that only samples one altitude slice. This suggests that distributing the points in the horizontal plane is more important than distributing them in the vertical direction for this particular environment and visibility volume. The sensor model creates visibility volumes with the most horizontal variation at the bottom of the shape as seen in Fig.~\ref{fig:solutionsamples}; therefore, sampling the visibility volumes at the bottom is the best way to produce samples with the greatest  horizontal variation.

\begin{figure}[ht]
  \centering
  \includegraphics[width = 0.9\textwidth]{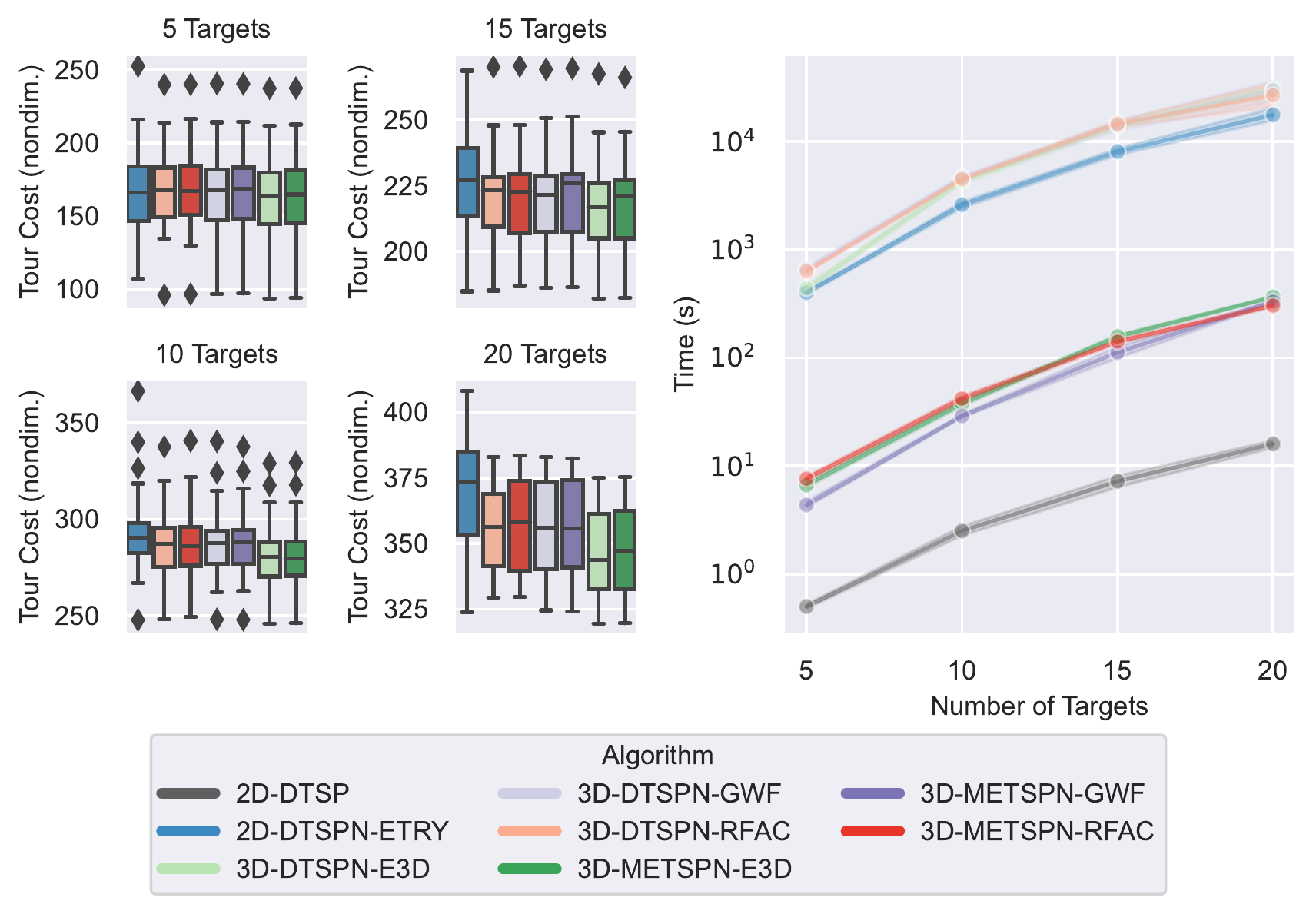}
  \caption{
    The box plots show the range of cost across all sets of target visibility volumes when the number of samples per target visibility is held at $n_{\rm pts} = 32$, the number of heading angles is held at $n_{\psi} = 8$ and the number of pitch angles is held at $n_\gamma = 1$.
    The vertical black bars show the upper and lower quartiles of the data while the colored sections show the middle quartiles.
    The black bar in the middle of the box plots is the median of the data set.
    The black diamonds are outliers. 
    The line graph shows the increase in computation time as the number of target visibility volumes increases on a $\log_{10}$ scale.
    The shaded region around each line shows the range of computation time.
  }
  \label{fig:samplingMethods}
\end{figure}

To isolate the effects of the different sampling methods, the results are examined for the case where the number of samples is held at $n_{\rm pts} =32$, the number of heading angles is held at $n_\psi = 8$, and the number of pitch angles is $n_\gamma=1$.
Box plots of those trials can be seen in Fig.~\ref{fig:samplingMethods}.
The medians of the 3D methods (black bar in the middle of the colored box) are lower than the medians of the 2D methods suggesting that the 3D methods are able to more consistently find lower-cost solutions.
The difference between medians of 2D and 3D methods grows as the number of target visibility volumes increases.
The range of solutions for the different methods, denoted by the vertical black bars, is large and suggests that the difference between the solutions produced by the 2D and 3D cases is variable and sensitive to the environment.
The time for each algorithm to execute on a single thread is shown in Fig.~\ref{fig:samplingMethods}.
It can be seen that the algorithms that only consider one point per region have lower execution times than the algorithms that consider neighborhoods.
The 2D ETRY method has a similar execution time to the 3D DTSPN methods.
However, the heuristic METSPN algorithm has a lower execution time compared to the other 3D methods because the graph that it creates is smaller and less computationally expensive.
The results suggest that for a large number of samples the METSPN algorithm outperforms the 3D DTSPN algorithms since it produces tours of similar cost but with a computation time that is approximately two orders of magnitude lower.

\section{Conclusion}
\label{sec:conclusion}
This paper studied the view planning problem of using a 3D Dubins airplane model to inspect points of interest in an urban environment in minimum time. Triangular meshes were used to compute approximate visibility volumes that correspond to locations where an unobstructed view of the target can be obtained while satisfying imaging and altitude constraints. The mesh-based approach for computing visibility volumes is flexible and can represent more complex geometries than have previously been considered. A range-based sensor model was assumed here, however mesh-based view planning can potentially support other sensor models, sensing modalities, and encode sensing performance characteristics. The 3D Dubins airplane model used in this work can, in some circumstances, produce more efficient inspection tours by exploiting altitude changes that are otherwise not possible with constant-altitude Dubins path tours. In cases where visibility volumes occupy disjoint altitude segments, the 3D algorithms provide a feasible solution where the 2D algorithms are not feasible. However, the pitch angle constraints of a Dubins airplane limit the change in altitude over a tour. Altitude changes are accompanied by an increase in path length and thus are only efficient when they greatly improve access to the visibility volume.

This work introduced a heuristic that computes edge costs by replacing the 3D Dubins path computation with a simpler lower bound and assigning heading and pitch angles based on the geometric relation of successive points in a tour. This strategy provides a similar tour cost to other 3D algorithms that use the exact 3D Dubins path planner for edge cost computation but with computation time reduced by two orders of magnitude.
Future work may consider the view planning problem in the presence of obstacles that must be avoided,
with target visibility volumes that overlap, and/or with uncertain moving targets to be inspected. 
\label{sec:conclusion}


\section*{Acknowledgments}
This work was supported by the William States Lee College of Engineering at the University of North Carolina at Charlotte through the  Multidisciplinary Team Initiation (MTI) Grant.
\bibliography{refs} 
\end{document}